\pdfoutput=1
\documentclass[letterpaper,12pt]{article}
\usepackage{tabularx}
\usepackage{amsmath,amssymb}  
\usepackage{graphicx}
\usepackage{physics}
\usepackage[margin=1in,letterpaper]{geometry}
\usepackage{cite}
\usepackage[final]{hyperref} 
\usepackage{comment}
\usepackage{fancyhdr}
\usepackage{authblk}

\hypersetup{colorlinks=true, linkcolor=blue, citecolor=magenta, filecolor=magenta, urlcolor=blue}

\begin{document}

\title{A domain wall and chiral edge currents in holographic chiral phase transitions}
\author[1]{Shuta Ishigaki}
\author[2]{Masataka Matsumoto}
\author[3,4]{Ryosuke Yoshii}
\affil[1]{Department of Physics, Shanghai University, Shanghai 200444, China}
\affil[2]{Wilczek Quantum Center, School of Physics and Astronomy, Shanghai Jiao Tong University, Shanghai 200240, China}
\affil[3]{Center for Liberal Arts and Sciences, Sanyo-Onoda City University, Yamaguchi 756-0884, Japan}
\affil[4]{International Institute for Sustainability with Knotted Chiral Meta Matter (WPI-SKCM2), Hiroshima University, Higashi-Hiroshima, Hiroshima 739-8526, Japan}
\date{}
\maketitle

\begin{abstract}
We investigate spatially inhomogeneous solutions in a top-down holographic model: the D3/D7 model which provides a holographic description of the chiral phase transition for a finite external magnetic field, chemical potential, and temperature.
We numerically find a domain wall (or kink) solution in the three dimensional space, which incorporates between the chiral symmetry broken phase at the spatial infinity, under the homogeneous sources.
Along with the inhomogeneity of the chiral condensate, the charge density is also spatially modulated.
The modulated charge density and finite magnetic field lead to the chiral edge current close to the domain wall.
We explore the dependences of those profiles on the chemical potential and temperature near the first and second order phase transition points.
Our results indicate that the inhomogeneous solutions we found are in good agreement with those obtained by the Ginzburg--Landau theory in the vicinity of the transition points.
\end{abstract}

\newpage
\tableofcontents
\section{Introduction}
For decades, the gauge/gravity duality (holography) has pioneered vast areas of physics.
The duality was originally proposed as the equivalence between a weakly coupled supergravity theory and a strongly coupled supersymmetric Yang-Mills (SYM) theory, the so-called Anti-de Sitter/conformal field theory (AdS/CFT) correspondence \cite{Maldacena:1998,Gubser:1998,Witten:1998}.
However, the correspondence has been generalized beyond the conformal case and extensively utilized to investigate strongly correlated many-body systems with the applications into condensed matter physics or QCD in mind. 

In the literature on applications of holography, there are generally two distinct models for investigating the dual descriptions of condensed matter physics or QCD: the top-down and bottom-up models.
In the top-down models, the dual field theory is rigorously given according to the AdS/CFT dictionary. 
For instance, the probe brane model, such as the D3/D7 model, is one class of the top-down models, where flavor degrees of freedom are introduced by adding the probe D7-brane in a stack of D3 branes. The corresponding theory is the ${\cal{N}}=4$ SYM theory with the ${\cal{N}}=2$ hypermultiplets \cite{Karch:2002sh}. Considering the probe brane model as a QCD-like model, for example, the chiral symmetry breaking \cite{Babington:2003vm,Evans:2010iy}, the phase transition between the meson/dissociation phases \cite{Mateos:2007vn}, meson spectra \cite{Kruczenski:2003be,Erdmenger:2007cm} were investigated. 
Moreover, the probe brane model was also employed under the motivation for condensed matter physics. For instance, the dual system can be considered as a many-body system consisting of charged particles and one can realize a conducting system by applying an external electric field \cite{Karch:2007pd}.

The bottom-up models are more phenomenological models, which contain a minimal field content to capture the essential dynamics in question.
A primary application of bottom-up models to condensed matter physics is the holographic superfluid model \cite{Hartnoll:2008kx,Hartnoll:2008vx}.
In the holographic superfluid model, the bulk U(1) gauge symmetry is spontaneously broken and the scalar field condensates in the AdS geometry.
Subsequently, spatially inhomogeneous solutions, such as single kink and vortex solutions, were constructed by assuming that the fields depend also on the spatial coordinate and their sources are homogeneous at the boundary \cite{Keranen:2009vi,Keranen:2009ss,Keranen:2009re,Albash:2009iq,Albash:2009ix,Montull:2009fe}.
Moreover, multiple kinks solutions and periodic kinks solutions were found in \cite{Lan:2017qxm,Matsumoto:2019ipg}.
While the order parameter in these inhomogeneous solutions is assumed to be real, more general inhomogeneous solutions, such as complex multiple kinks and twisted kink crystals, were also discovered by considering the complex scalar fields\cite{Matsumoto:2020mgi}.

Compared to the bottom-up model, little is known about inhomogeneous solutions in the top-down model.
A known example is in the D3/D7' model which describes (2+1)-dimensional Fermi-liquid like systems \cite{Bergman:2010gm}. In this model, the striped instability was observed by linear perturbation analysis \cite{Bergman:2011rf}. Subsequently, inhomogeneous solutions corresponding to a spin and charge density wave states were found by numerically solving partial differential equations \cite{Jokela:2014dba}.
In the holographic QCD model, namely top-down Sakai-Sugimoto model \cite{Sakai:2004cn}, similar instabilities have been reported \cite{Ooguri:2010xs,Chuang:2010ku,Bayona:2011ab,Ballon-Bayona:2012qnu}.
In the D3/D7 model as well, the dynamical instabilities leading to spatially inhomogeneous states have been discussed, for example, a chiral helix phase \cite{Kharzeev:2011rw} and a current filament in the steady state \cite{Ishigaki:2021vyv}.
To our knowledge, however, inhomogeneous solutions, such as a kink solution, have not been found in the D3/D7 model.

Apart from holography, the spatially inhomogeneous solutions are universally observed in various models, whereas whether these are physically realized or not depends on each specific condition.
In condensed matter physics, it has been known that strongly correlated superconductors exhibit several inhomogeneous phases such as the Larkin--Ovchinnikov--Fulde--Ferrel (LOFF) phase \cite{Larkin,Fulde}. 
In a Bardeen--Cooper--Schrieffer (BCS) superconductor, the LOFF phase is expected to be stabilized by applying a large magnetic field at low temperature. The effects of the magnetic field appears in two manners; the one is the Aharonov-Bohm phase which results in the phase modulation of the order parameter and the other is the Zeeman splitting which induces the excess spins and thus results in the formation of nodes of the order parameter \cite{Quan,Yoshii:2015}. In experiment, the indirect evidences for the existence of the LOFF phase have been reported in the heavy fermion superconductor CeCoIn$_{5}$ \cite{Radovan2003magnetic,Bianchi2003possible}, the organic superconductor $\kappa\text{-}(\text{BEDT-TTF})_{2} \text{Cu(NCS)}_{2}$ \cite{Cho2009upper,Coniglio2011superconducting,Agosta2012experimental,Agosta2017}, and also in the cold atom system \cite{Liao2010}.
On the other hand, it is well-established that the rich phase structure of QCD incorporates the spatially inhomogeneous phases in which the chiral condensate is modulated (see reviews \cite{Casalbuoni2004inhomogeneous,Buballa2015inhomogeneous}). The relevant solutions has been investigated in the (1+1)-dimensional Nambu--Jona-Lasinio model or the chiral Gross--Neveu model \cite{Basar2008self,bacsar2008twisted,Basar2009inhomogeneous}.  Regarding inhomogeneous condensations, the connection to the non-linear sigma model has also been discussed (see review \cite{Yoshii:2019yln} and references therein). The subsequent study with the generalized Ginzburg--Landau expansion turned out that the phase with spatially inhomogeneous chiral condensates is energetically preferred \cite{Nickel2009many}.
In this way, spatially inhomogeneous solutions are of significant interest both in condensed matter physics and in QCD.

\begin{figure}[tbp]
    \centering
    \includegraphics[width=7.5cm]{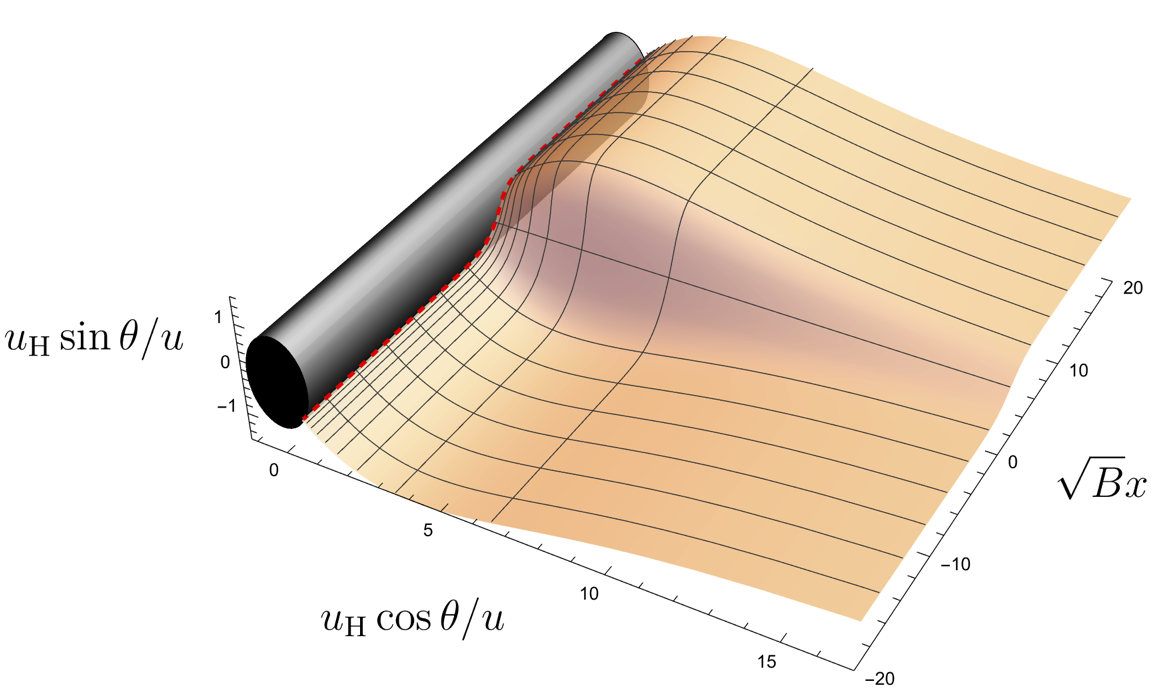}
    \includegraphics[width=7.5cm]{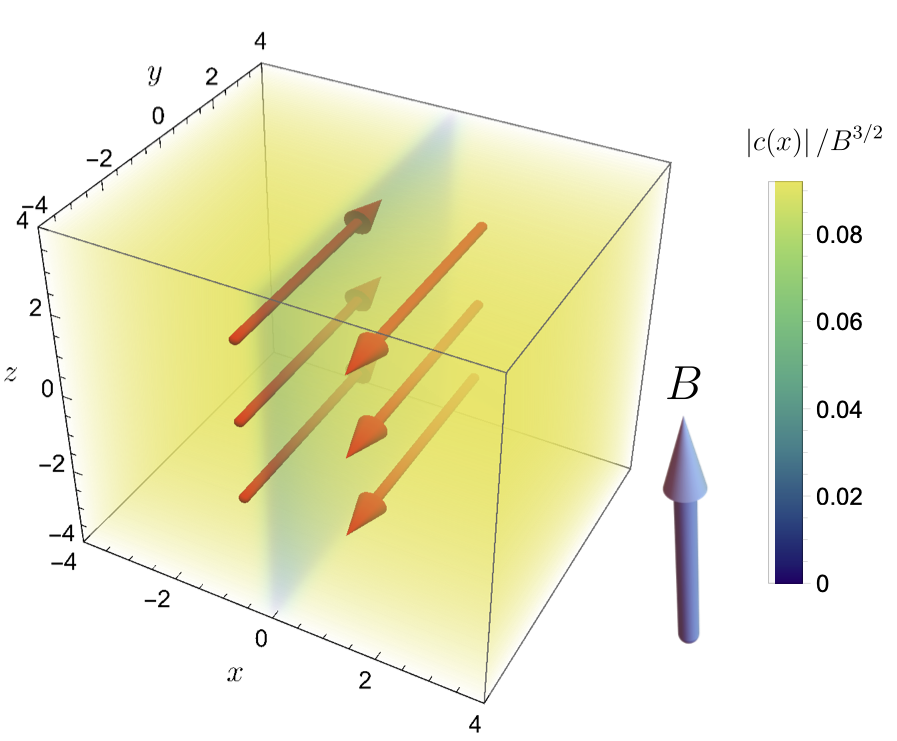}
    \caption{The left panel shows the spatially inhomogeneous profiles of the probe D7-brane embedding we found in this paper. The embedding reach the black hole horizon (red dashed line) and approach zero at the AdS boundary ($u=0$). The right panel shows the domain wall structure in the three dimensional space $(x,y,z)$. The colors indicate the amplitude of the chiral condensate $c(x)$ normalized by the magnetic field $B$ along $z$-direction. Here, $c(x)$ is odd function of $x$ and thus the $\pi$ phase shift of the chiral condensate takes place at $x=0$. The red arrows schematically denote the current density along $y$-direction beside the domain wall.}
    \label{fig:dwj}
\end{figure}

Motivated by these, in this paper we study spatially inhomogeneous solutions, especially a real kink solution, associated with the chiral symmetry breaking in the D3/D7 model. 
These inhomogeneous solutions correspond to the inhomogeneous profiles of the probe D7 brane (and gauge fields on the D7 brane) as shown in the left panel of figure \ref{fig:dwj}.
Interestingly, we find that the {\it chiral edge current} induced by the magnetic field is accompanied with an emergence of kink.
The right panel of figure \ref{fig:dwj} shows the dual picture of the kink (central blue region) and chiral edge current (red arrows) in the dual three dimensional space.

The main differences between our findings and previous studies of a kink solution in the bottom-up model \cite{Keranen:2009re,Keranen:2009ss,Keranen:2009vi,Albash:2009iq,Albash:2009ix,Montull:2009fe} are the following: (1) Since our model is the top-down D3/D7 model, the dual description is clear; we find the single real kink solution with the chiral condensate as the order parameter in the ${\cal{N}}=4$ large-$N_{c}$ SYM theory with ${\cal{N}}=2$ fundamental hypermultiplets.
(2) To find the spontaneous chiral symmetry breaking even in the homogeneous setup, the external magnetic field is necessary \cite{Evans:2010iy}. It is known that the magnetic field in the D3/D7 model plays a role of conformal symmetry breaking in the same fashion as $\Lambda_{\rm QCD}$ in QCD. We find that it also induces the chiral edge current accompanied with the inhomogeneity.
(3) The chiral symmetry breaking in the D3/D7 model occurs by controlling two independent parameters, namely temperature and charge density (scaled by the magnetic field). Hence, both of first and second order phase transitions are observed in the phase diagram, whereas the holographic superfluid model undergoes only the second order phase transition. In our study, we investigate profiles and properties of inhomogeneous solutions close to first and second order phase transition points.

This paper is organized as follows.
In section \ref{sec:setup}, we briefly review the chiral symmetry breaking in the D3/D7 model and explain our setup to find inhomogeneous solutions.
Before we proceed to study the inhomogeneous solutions by solving the nonlinear equations, we briefly discuss the inhomogeneous scale obtained from the linear perturbation analysis in section \ref{sec:static}.
In section \ref{sec:inhomo}, we show the inhomogeneous solutions for several temperatures and chemical potentials obtained from the nonlinear equations. 
We study the deformation of profiles and amplitude of chiral edge current with respect to the parameters. 
We also compare them to the solutions derived from the Ginzburg--Landau theory.
At last, the analysis of the thermodynamic stability is performed by computing the thermodynamic potential.
Section \ref{sec:discussion} is devoted to the conclusions and discussions.

\section{Holographic setup} \label{sec:setup}
In our study, we employ the top-down D3/D7 model in order to investigate inhomogeneous chiral condensates. In the spatially homogeneous setup, the D3/D7 model provides the holographic description of the chiral symmetry breaking in the presence of the chemical potential and magnetic field at finite temperature \cite{Evans:2010iy}.
In this section, we explain our setup to find inhomogeneous solutions and briefly review the phase diagram of the chiral symmetry breaking with the homogeneous chiral condensate.

The background metric is the AdS-Schwarzschild black hole times $S^{5}$:
\begin{equation}
    \dd s^{2} =\frac{1}{u^{2}} \left[ -f(u) \dd t^{2} + \frac{1}{f(u)} \dd u^{2} +\dd \vec{x}^{2} \right] + \dd \Omega_{5}^{2},
\end{equation}
where we set the radius of AdS and $S^{5}$ part to one and
\begin{equation}
    f(u)=1-\frac{u^{4}}{u_{\rm H}^{4}},
\end{equation}
with the black hole horizon $u_{\rm H}$. Here $t$ and $\vec{x}=(x,y,z)$ are the four dimensional spacetime coordinates and $u$ is the radial direction of the AdS geometry.
The Hawking temperature $T = 1/ (\pi u_{\rm H})$ corresponds to the temperature of the heat bath in the boundary theory.
The metric of the $S^{5}$ part is given by
\begin{equation}
    \dd \Omega_{5}^{2} = \dd \theta^{2} + \sin^{2}\theta \dd \psi^{2} + \cos^{2}\theta \dd \Omega_{3}^{2}.
    \label{eq:Omega5}
\end{equation}
On this background, the dynamics of the probe D7-brane are governed by the Dirac--Born--Infeld (DBI) action with Wess--Zumino (WZ) term.
The DBI action is given by
\begin{equation}
    S_{\rm DBI} = -T_{D7} \int d^{8}\xi \sqrt{-\det(g_{ab}+ 2\pi \alpha'F_{ab})},
    \label{eq:DBI}
\end{equation}
where $T_{D7} = (2\pi)^{-7} g_{\rm s}^{-1} \alpha'^{-4}$ represents the tension of the probe D7-brane.
The induced metric $g_{ab}$ in the eight dimensional spacetime with $a,b=0,\cdots, 7$ is defined by
\begin{equation}
    g_{ab} = \partial_{a}X^{M} \partial_{b}X^{N} G_{MN},
\end{equation}
where $X^{M}$ denotes the target space coordinate and $G_{MN}$ is the background metric in the ten-dimensional spacetime with $M,N = 0,\cdots,9$. 
The field strength of the U(1) gauge field on the D7-brane is given by $2\pi \alpha' F_{ab}=\partial_{a}A_{b}-\partial_{b}A_{a}$.
For our purpose, the ansatz for the fields is
\begin{eqnarray}
    \theta &=& \theta(u,x), \\
    A_t &=& a_{t} (u,x), \\
    A_{y} &=& Bx + a_{y}(u,x),
\end{eqnarray}
where $B$ is related to a constant magnetic field along $z$-coordinate. 
Without loss of generality, we choose the gauge of $A_{u}=0$ and assume $\psi = 0$.
Under this ansatz, we ignore the other gauge field components ($A_{x}, A_{z}$) that are decoupled to the other fields.
We can also ignore the WZ term because it has no contribution in this setup.
From the DBI action (\ref{eq:DBI}), we obtain the nonlinear partial differential equations for the fields ($\theta,a_{t},a_{y}$).
We will not present the explicit forms of them because they are highly complicated and not very illuminating.

The boundary conditions at each boundary are imposed as follows.
At the horizon $u=u_{\rm H}$, we impose the regularity conditions for the fields, which are derived from the equations of motion.
In the vicinity of the AdS boundary $u=0$, the asymptotic behaviors of the field are written as
\begin{eqnarray}
    \frac{\sin\theta(u,x)}{u} &=& m + c(x) u^{2} + \cdots, \label{eq:asym1}\\
    a_{t}(u,x) &=& \mu - \frac{\rho(x)}{2} u^{2} + \cdots, \label{eq:asym2}\\
    a_{y}(u,x) &=& b + \frac{J_{y}(x)}{2} u^{2} + \cdots,
    \label{eq:asym3}
\end{eqnarray}
where $m$, $\mu$, and $b$ are related to the sources for the dual operators \cite{Kobayashi:2006sb} as
\begin{equation}
    m_{q} = \frac{\lambda^{1/2}}{2\pi}m, \quad \mu_{q} = \frac{\lambda^{1/2}}{2\pi} \mu,
\end{equation}
where $m_{q}$ and $\mu_{q}$ are the quark mass and chemical potential, respectively.
Here, we used the relation $4\pi g_{s}N_{c} \alpha'^{2} = 1$ with the 't~Hooft coupling given by $\lambda = g_{\rm YM}^{2}N_{c} =4\pi g_{s} N_{c} $.
We assume that the source for $a_{y}$ vanishes: $b=0$.
The dual operators $c(x)$, $\rho(x)$, and $J_{y}(x)$, respectively, correspond to the chiral condensate, charge density, and current density along $y$-direction via
\begin{equation}
    \left< \bar{q}q\right> = 2\frac{N_{c}}{(2\pi)^{3}}\lambda^{1/2}c(x), \quad \left<\bar{q}\gamma^{0}q \right> = \frac{N_{c}}{(2\pi)^{3}}\lambda^{1/2}\rho(x), \quad \left<\bar{q}\gamma^{y}q \right> = \frac{N_{c}}{(2\pi)^{3}}\lambda^{1/2}J_{y}(x).
\end{equation}
Hereafter, for simplicity, we regard the geometric parameters $m,\mu,c,\rho$, and $J_{y}$ as the corresponding sources and operators in the dual field theory.
Since we are interested in the spontaneous breaking of the chiral symmetry and translation symmetry, we set $m_q=0$ ($m=0$) and assume that $\mu_q$ ($\mu$ also) is a spatially homogeneous constant. 
Note that we assume these operators to be spatially inhomogeneous with respect to the homogeneous sources.

Additionally, we require to impose the boundary conditions at the spatial infinity $x=\pm \infty$. 
At these points, we impose the Neumann boundary conditions so that the fields approach the homogeneous profiles.
For technical reasons, we fix the size of the system with respect to $x$-coordinate by introducing a cutoff at $x=\pm l /2$ and impose the Neumann boundary conditions there.

\begin{figure}[t!bp]
    \centering
    \includegraphics[width=10cm]{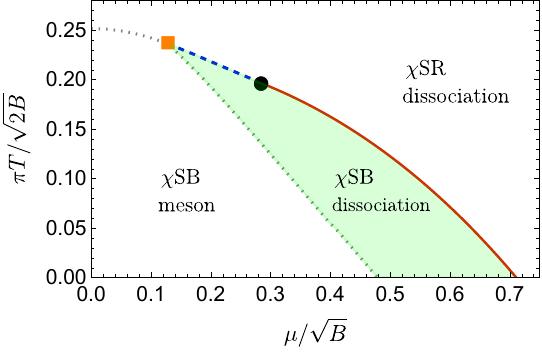}
    \caption{The phase diagram in the temperature and chemical potential plane with massless quarks. The red and blue dashed curves denote the second and first order phase transition lines of the chiral transition, respectively. The black dot denotes the tricritical point. The gray dotted curve denotes the second order phase transition line of the meson/dissociation phase transition. The orange square denotes the point where those two transitions merge. The phase diagram was originally shown in \cite{Evans:2010iy}. In the $\chi$SB phase (light green shaded region), we find the inhomogeneous solutions.}
    \label{fig:phase}
\end{figure}

Before we proceed to discuss the inhomogeneous chiral condensates, we should mention the phase diagram of the chiral transitions in the spatially homogeneous setup.
In \cite{Evans:2010iy}, the spatially homogeneous chiral condensates were found and the phase diagram with respect to the temperature and chemical potential (normalized by the magnetic field) were presented.
Figure \ref{fig:phase} shows the phase diagram of the D3/D7 model in the homogeneous setup.
As shown, the chiral phase transition (red solid and blue dashed curves) and the meson/dissociation phase transition (gray dotted curve) are observed.
In our study, we focus on the chiral transition between the chiral symmetry restored ($\chi$SR) and broken ($\chi$SB) phase because we are interested in a kink-like inhomogeneous solution which connects the $\chi$SB phases via the $\chi$SR phase.
Hence, we focus on the parameter region around the phase boundary between $\chi$SB dissociation and $\chi$SR dissociation phase in figure \ref{fig:phase}. 

\section{Inhomogeneous static perturbation} \label{sec:static}
\begin{figure}[tbp]
\centering
    \includegraphics[width=8.3cm]{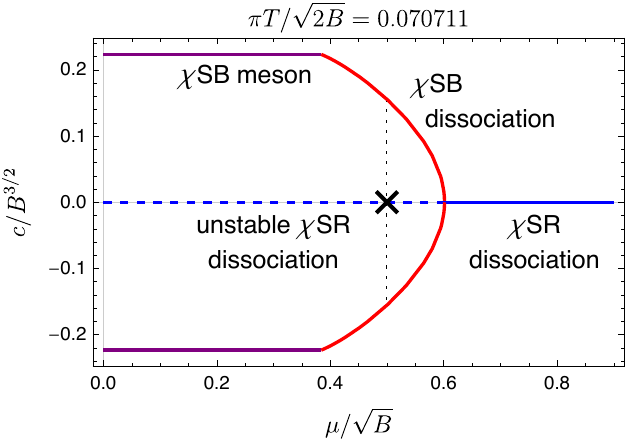}
    \caption{
    Chiral condensate $c/B^{3/2}$ vs chemical potential $\mu/\sqrt{B}$ of several phases at $\pi T/\sqrt{2B}=0.070711$.
    The cross denotes the unstable $\chi$SR solution at $\mu/\sqrt{B}=0.5$, which corresponds to Fig.~\ref{fig:mu-ksp}.
    }
    \label{fig:mu-c}
\end{figure}

Prior to solving the nonlinear equations, we shall study the inhomogeneous length scale indicated by the linear perturbation analysis on the solutions in the $\chi$SR phase. 
According to the linear perturbation analysis, when the $\chi$SB solution realizes, the $\chi$SR solution described by the flat embedding becomes dynamically unstable in the homogeneous setup.
Figure \ref{fig:mu-c} shows the $\mu$--$c$ curves of various solutions. 
In the unstable $\chi$SR vacuum, one can find a tachyonic mode for the perturbation of the scalar field.
The tachyonic mode has a purely-imaginary frequency $i\omega$ with $\omega>0$, which implies the growing of the small perturbation in time and thus results in the instability. 
If we consider the spatial modulation of the perturbation by the wavenumber $k_x$, the growth rate can be suppressed.
At a specific wavenumber $k_x=k_{\text{sp}}$, in particular, the mode becomes static with zero frequency.
We refer to it as the inhomogeneous static perturbation.
The presence of the static perturbation at finite wavenumber suggests the presence of the inhomogeneous solution. 
Though the nonlinear analysis is necessary to conclude whether such an inhomogeneous solution becomes stable or not, the linear perturbation analysis is helpful for a clear understanding. 
Here, we show some results for the inhomogeneous static perturbation.

We are interested in the linear perturbation of the scalar field corresponding to the direction of the $\chi$SB. 
Now, we consider the plane-wave ansatz
\begin{equation}
    \theta \to \vartheta(u) e^{-i(\omega t - k_x x)},
\end{equation}
where $\vartheta(u)$ is a small perturbation field.
Since this is decoupled from the other perturbations under the current assumption, the linearized equation of motion is simply given by
\begin{equation}
    \left[
        \frac{3 B^2 u^4-k_x^2 u^2+3}{
            u^2 f (B^2 u^4+\rho^2 u^6+1)
        }
        +\frac{\omega^2}{f^2}
    \right]\vartheta(u)
    +
    \left[
        \frac{f'}{f}
        -\frac{B^2 u^4+3}{B^2 u^5+\rho^2 u^7+u}
    \right]\vartheta'(u)
    +\vartheta''(u) = 0.
    \label{eq:eom_scalar_perturbation}
\end{equation}
We impose the ingoing-wave boundary condition at $u=u_{\rm H}$.
The asymptotic expansion is given by $\vartheta(u) = \theta_{0} u + \theta_{2} u^{3} + \cdots$.
To obtain the mode solution, we also impose that the non-normalizable term vanishes, i.e., $\theta_{0}=0$.

\begin{figure}[tbp]
\centering
    \includegraphics[width=7cm]{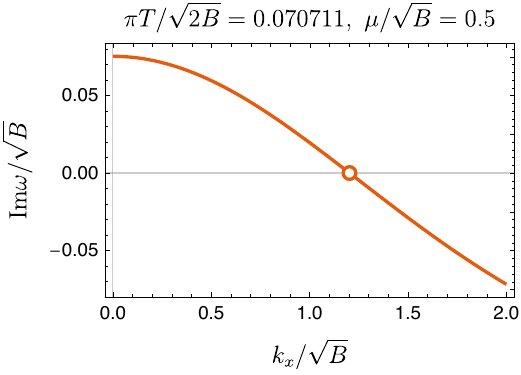}
    \includegraphics[width=7cm]{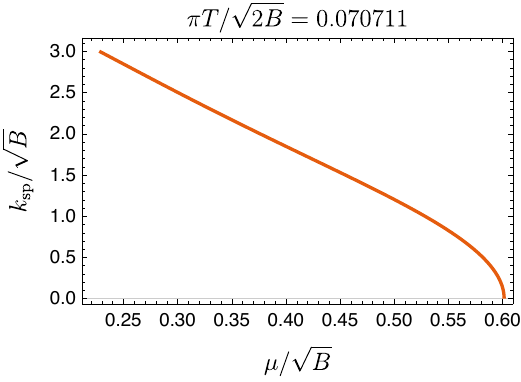}
    \caption{
    (left) Dispersion relation of the tachyonic mode in the $\chi$SR phase.
    (right) Wavenumber of the static perturbation as a function of $\mu$.
    }
    \label{fig:mu-ksp}
\end{figure}

The left panel of figure \ref{fig:mu-ksp} shows the $k_x$-dependence of the tachyonic mode-frequency, i.e., the dispersion relation, for $(\pi T/\sqrt{2B},\mu/\sqrt{B}) = (0.070711,0.5)$.
At $k_x=0$, the positive imaginary part implies the instability.
Remark that the mode frequency is purely imaginary. 
As $k_x$ increase, $\mathrm{Im}\omega$ decreases and reaches zero at a specific wavelength $k_x = k_{\text{sp}}$, i.e., the inhomogeneous static perturbation is realized. 
The result implies that the long-wavelength modes can grow at the unstable vacuum, whose wavelength are longer than $1/k_{\text{sp}}$.
Though the linear analysis cannot predict what the final state will be, we can naively expect that $1/k_{\text{sp}}$ corresponds to the length scale emerged in the final state.

The right panel of figure \ref{fig:mu-ksp} shows $k_{\text{sp}}$ as a function of $\mu$ with the fixed temperature (by taking a scale as $B$).
At $\pi T = 0.070711\sqrt{2B}$, the second order phase transition point locates at $\mu=\mu_c = 0.60146 \sqrt{B}$.
From the fitting near the phase transition point at this temperature, $k_{\text{sp}}$ is found to be 
\begin{equation}
    \frac{k_{\text{sp}}}{\mu_c} \approx
    4.62 \sqrt{1 - \frac{\mu}{\mu_c}}.
    \label{eq:staticpert}
\end{equation}
In the later section, we show that this result obtained from the linear perturbation analysis, especially the $\mu$-dependence of the inhomogeneous length scale (\ref{eq:staticpert}), agrees with the result obtained from the nonlinear analysis.

\section{Kink solutions} \label{sec:inhomo}
In this section, we study the spatially inhomogeneous solutions obtained by numerically solving the nonlinear partial differential equations.
The details of the numerical method are discussed in Appendix \ref{sec:numeric}.
The size of the system is fixed to $l=8$ in our numerical calculations.

\subsection{Kink and chiral edge current} \label{Kink-chiraledge}
\begin{figure}[tbp]
\centering
    \includegraphics[width=8cm]{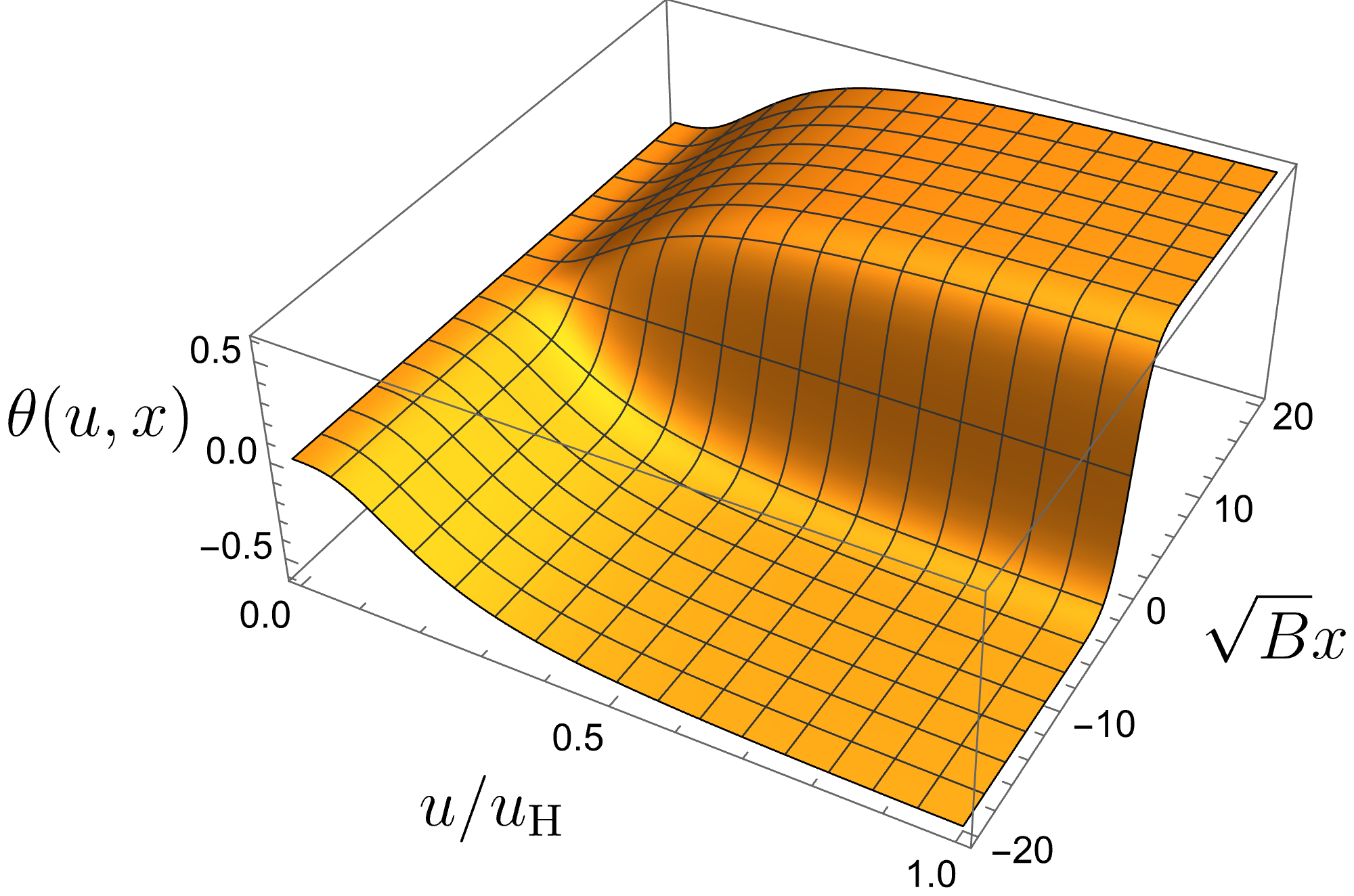}
    \includegraphics[width=8cm]{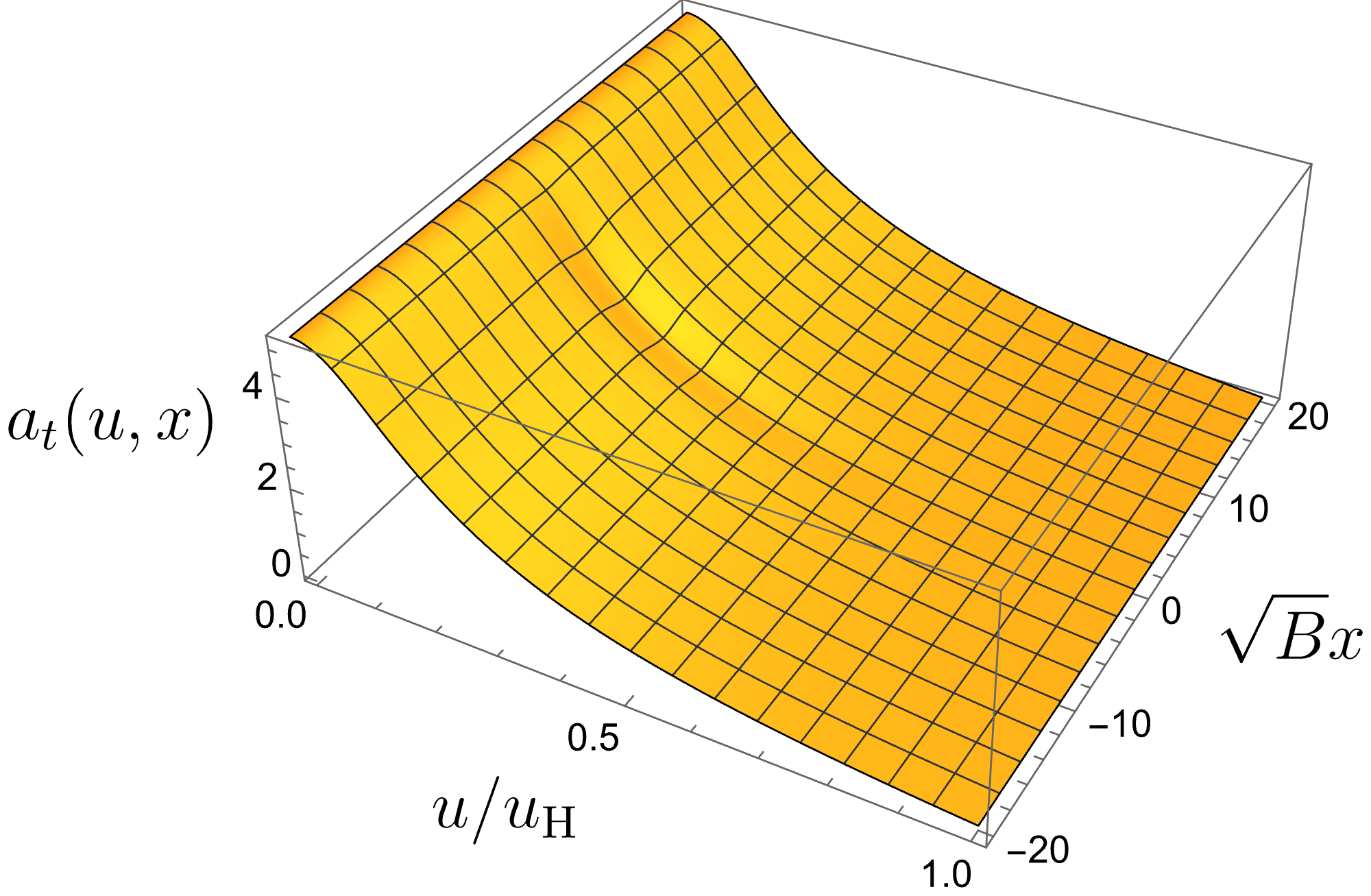}
    \includegraphics[width=8cm]{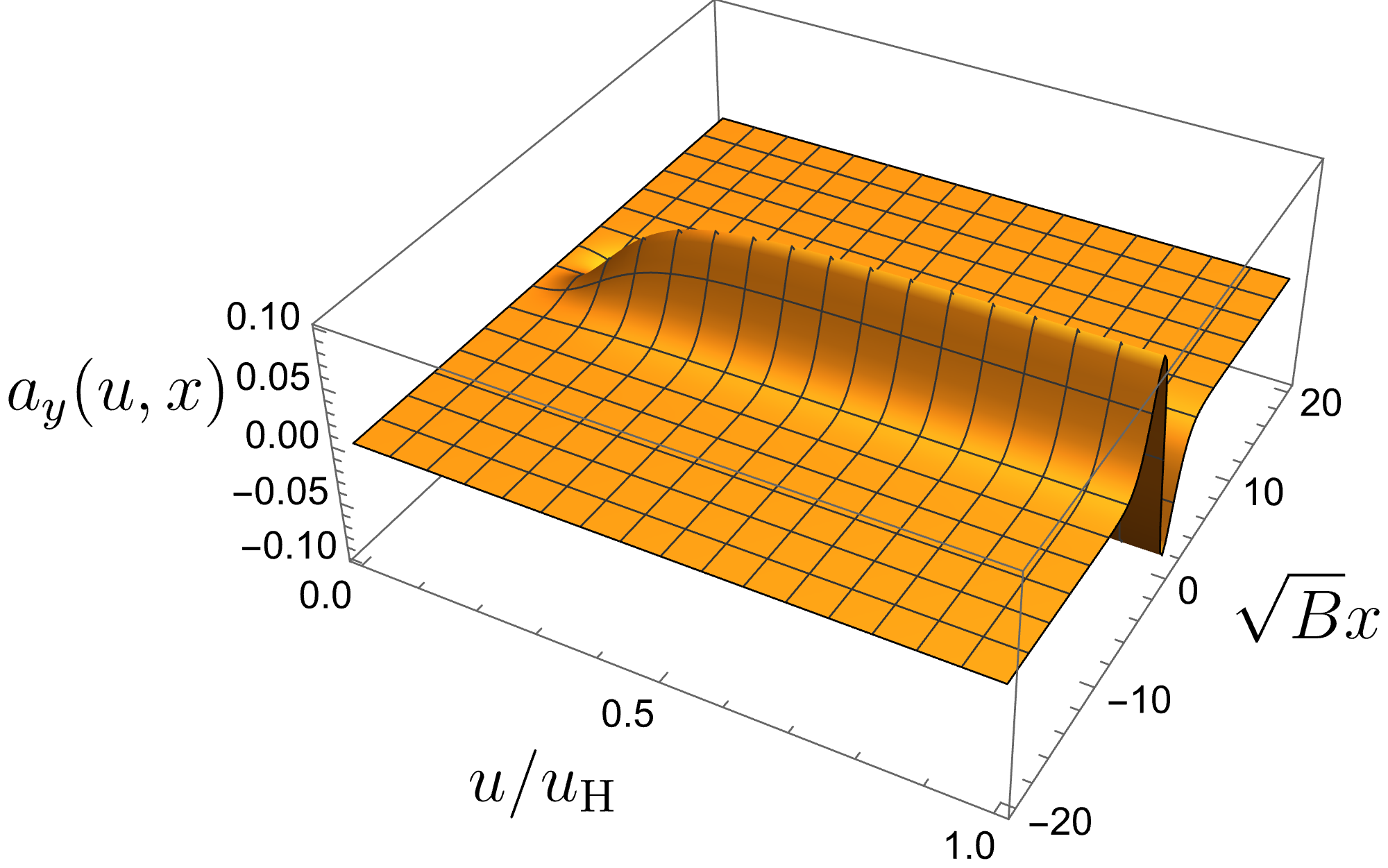}
    \caption{The typical profiles of $\theta(u,x)$, $a_{t}(u,x)$, and $a_{y}(u,x)$ for a kink solution for $\mu/\sqrt{B} =0.55$ and $\pi T/\sqrt{2B} = 0.075$.}
    \label{fig:3d}
\end{figure}

\begin{figure}[tbp]
    \includegraphics[trim = 15 10 0 0, scale=0.55, clip]{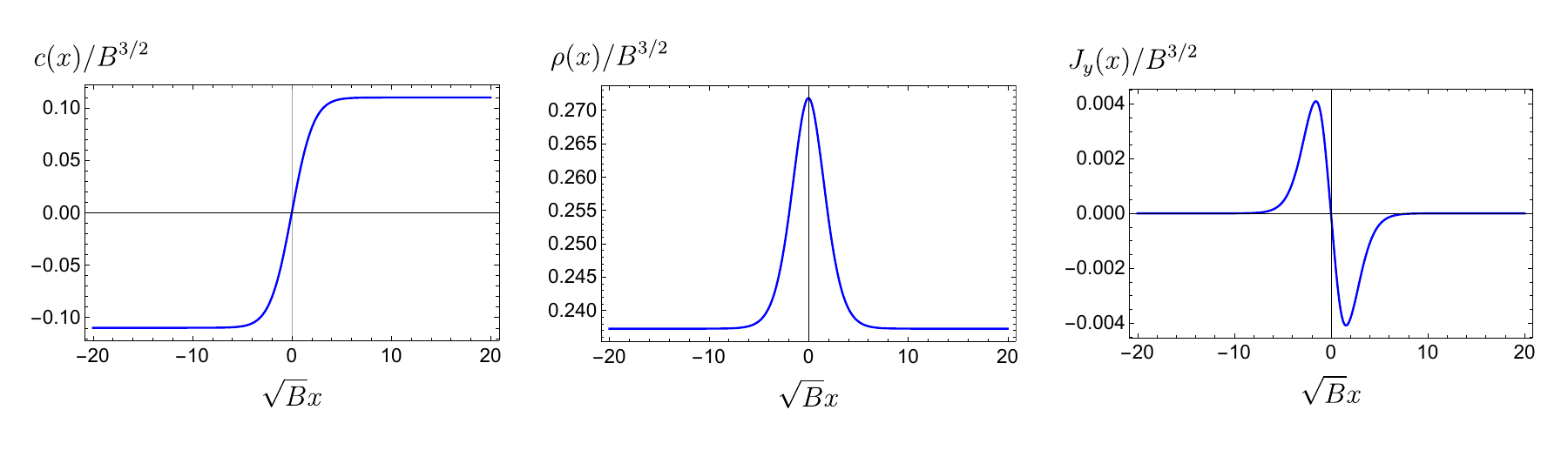}
    \caption{The profiles of the dual operators $c(x)$, $\rho(x)$, and $J_{y}(x)$ obtained from the kink solution.}
    \label{fig:bdry}
\end{figure}

We numerically find the spatially inhomogeneous solutions with a single kink.
Figure \ref{fig:3d} shows the typical profiles of $\theta(u,x)$, $a_{t}(u,x)$, and $a_{y}(u,x)$ for $\mu/\sqrt{B} =0.55$ and $\pi T/\sqrt{2B} = 0.075$.
From the asymptotic behaviors near the AdS boundary (\ref{eq:asym1})-(\ref{eq:asym3}), we can read off the profiles of the dual operators for the kink solution.
Figure \ref{fig:bdry} shows the profiles of $c(x)$, $\rho(x)$, and $J_{y}(x)$ normalized by $B$.
The profile of $c(x)$ (left panel) shows the kink located at the origin ($x=0$); the chiral condensate is homogeneous with the phase $0$ and $\pi$ at $x\to\infty$ and $x\to -\infty$, respectively, and those are connected at $x=0$ where the order parameter vanishes. 
Since we assume the phase of the order parameter to be zero or $\pi$ everywhere, the inhomogeneous solutions can be considered as single real-kink solutions. 
We observe that the charge density becomes locally large at the position of the kink (middle panel). 
The behavior is consistent with the result in the homogeneous setup; the charge density is always large in the chiral symmetry restored phase ($c=0$) with the chemical potential fixed compared to the broken phase ($c\neq0$) \cite{Evans:2010iy}. 
Physically, this behaviour comes from the fact that the condensates acts as the potential for charges and thus the charges tend to accumulate to the place where the amplitude of the condensate is small. As a consequence, the locally restored $\chi$SB (such as kink solutions) becomes more favored compared with the homogeneous $\chi$SB in some situations.

An interesting feature of this kink solution is that the finite electric current, $J_{y}(x)$, is induced along the $y$-direction (right panel).
This current can be interpreted as the {\it chiral edge current} for the following reason.
The dual system considered here has a finite charge density and magnetic field along the $z$-direction at finite temperature.
Then, the charged particles are supposed to move in the plane perpendicular to the magnetic field by the Lorentz force. 
Notice that the charged particles may be effectively massive because of the finite chiral condensate while the bare mass is zero.
In the homogenous $(3+1)$d bulk, we cannot observe the current because each cyclotron motion of the particles nearby contributes to the current with the opposite directions and is microscopically canceled out.
At the edge of the system, however, the current can be observed, which is referred to as the chiral edge current.
In our setup, the kink plays a role of the edge of the two separated bulk systems, namely $-l/2\leq x< 0$ and $0<x\leq l/2$.
The gradient of the charge density is essential to observe the chiral edge current because the cancelation of the cyclotron motions is unbalanced where the charge density is varied.
Remarkably, the chiral edge current is dissipationless because there is no Joule heating: $\vec{J}\cdot\vec{E} = 0$.
We numerically observe that the profile of $J_{y}(x)$ is proportional to $\partial_x \rho(x)$.
Formally, we can write
\begin{equation}
    J_y(x) \simeq \gamma(B,\mu,T) \partial_x \rho(x),
    \label{eq:chiral_edge_current}
\end{equation}
where $\gamma$ is a positive constant depending on the parameters $B, \mu$ and $T$.
From the scaling analysis, $\gamma$ must be a quantity with a (scaling) dimension $-1$.
In the case of figure \ref{fig:bdry}, the coefficient is obtained as $\sqrt{B} \gamma \approx 0.3426$ by fitting.
Note that the sign of $J_y$ must be flipped under the time-reversal transformation $\mathcal{T}$.
Since $B$ is only the $\mathcal{T}$-odd constant in our setup, $\gamma$ must be $\mathcal{T}$-odd.
It implies that $\gamma$ must be an odd function of $B$.

\subsection{Chemical potential/temperature dependence}
\begin{figure}[tbp]
    \centering
    \includegraphics[width=7cm]{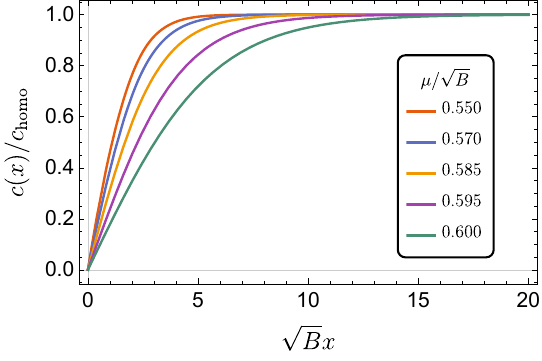}
    \includegraphics[width=7cm]{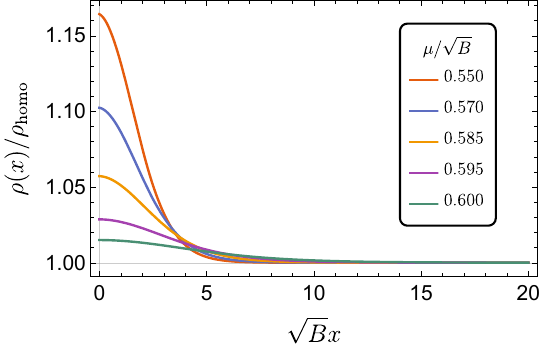}
    \caption{The profiles of $c(x)$ (left) and $\rho(x)$ (right), normalized by the homogeneous values $c_{\rm homo}$ and $\rho_{\rm homo}$, with $\pi T/\sqrt{2B} = 0.070711$ fixed.}
    \label{fig:cmrm}
\end{figure}

Now we study the behaviors of the kink solutions with respect to the chemical potential with temperature fixed.
Figure \ref{fig:cmrm} shows the half profiles of $c(x)$ and $\rho(x)$ for several values of the chemical potential with $\pi T/\sqrt{2B} = 0.070711$ fixed. 
Here, each profile are normalized by the values obtained from the homogeneous ($\chi$SB) solutions: $c_{\rm homo}$ and $\rho_{\rm homo}$, which are coincident with the values at the spatial boundary.

\begin{figure}[tbp]
    \centering
    \includegraphics[width=7cm]{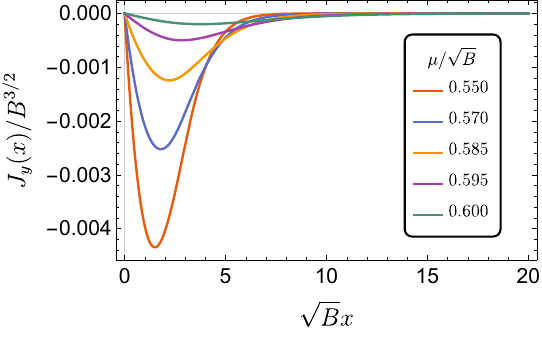}
    \includegraphics[width=7cm]{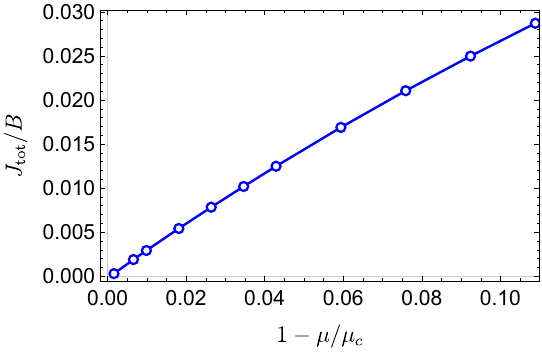}
    \caption{The profiles of $J_{y}(x)$ for several chemical potentials (left) and the chemical potential dependence of $J_{\rm tot}$ (right) with $\pi T/\sqrt{2B} = 0.070711$ fixed.}
    \label{fig:jym}
\end{figure}

It is shown that the width of the kink and the peak of the charge density becomes broader as the chemical potential increases.
For a large chemical potential the inhomogeneous solutions approach the homogeneous ones in the chiral symmetry restored phase. 
In the present case with $\pi T/\sqrt{2B} =0.070711$, the inhomogeneous solutions are not found above the second order phase transition point, $\mu_{c}/\sqrt{B}\approx 0.601$.
We also show the chemical potential dependence of $J_{y}(x)$ profiles and the total edge current in figure \ref{fig:jym}. 
Note that they are normalized by $B$ because $J_{y}(x)$ approach zero at the spatial boundary, namely $J_{y,\rm homo}=0$. 
Here, we define the total chiral edge current as 
\begin{equation}
    J_{\rm tot} \equiv - \int_{-l/2}^{l/2} dx\ {\rm sgn}(x)J_{y}(x).
\end{equation}
In the present setup, we find that the usual total current $J_{\rm tot}^\prime \equiv \int_{-l/2}^{l/2} dx J_{y}(x)$ becomes zero within the computational accuracy, indicating that there is no net effect from the chiral edge current. 
Thus we focus on the chiral edge current in the following. 
We find that the peak amplitude of the chiral edge current becomes large and the peak position approach the kink position as the chemical potential decreases (see figure \ref{fig:jym}). 
This increase of the total chiral edge current reflects the fact that the decrease of the chemical potential corresponds to leaving away from the second order phase transition point.
These behaviors are consistent with the fact that the chiral edge current is roughly given by the spatial derivative of the charge density as discussed above. 

\begin{figure}[tbp]
    \centering
    \includegraphics[width=7cm]{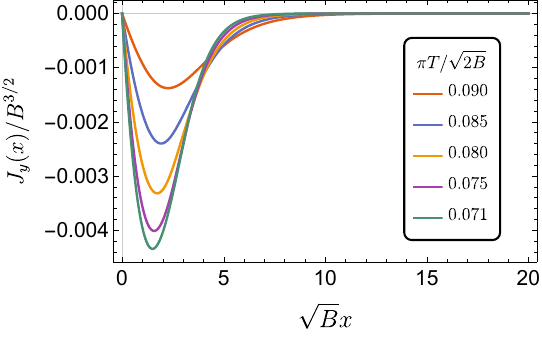}
    \includegraphics[width=7cm]{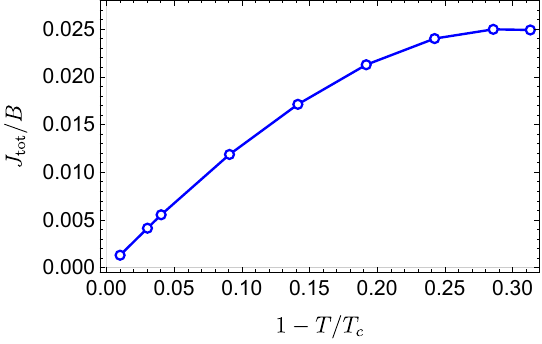}
    \caption{The profiles of $J_{y}(x)$ for several temperatures (left) and the temperature dependence of $J_{\rm tot}$ (right) with $\mu/\sqrt{B} = 0.55$ fixed.}
    \label{fig:jyT}
\end{figure}

We also study the temperature dependencies of the profiles of $c(x)$, $\rho(x)$, and $J_{y}(x)$.
We find that they are qualitatively similar to the chemical potential dependence; the profiles approach the homogeneous ones in the chiral symmetry restored phase as the temperature increase and approach the second order phase transition point $T_{c}$. 
For readability, we show only the temperature dependence of $J_{y}(x)$ profiles and $J_{\rm tot}$ in figure \ref{fig:jyT}. 
Interestingly, we find that the total chiral edge current does not monotonically increase with respect to temperature as shown in the right panel of figure \ref{fig:jyT}; it appears to have the maximum value at a specific value of temperature.

\subsection{Comparison to the 4th order Ginzburg--Landau theory} \label{sec:4thGL}
In this subsection, we compare our kink solutions with the results from the phenomenological effective theory: the Ginzburg--Landau (GL) theory.
A similar discussion was given in the context of dark solitons in holographic superfluids \cite{Keranen:2009vi,Keranen:2009ss}.
In the vicinity of the second order phase transition point, we assume that the thermodynamic potential $\Omega$ as a functional of a complex scalar field $\psi$ as 
\begin{equation}\label{eq:4th-GL}
    \Omega[\psi] =\int \dd^{3} x \left\{
    \frac{1}{2}\big| \vec{\nabla} \psi\big|^{2} +V(|\psi|)
    \right\},
    \quad
    V(\psi)= \frac{r_{0}}{2}\left|\psi\right|^{2} + \frac{g}{4} \left| \psi \right|^{4} + V_0,
\end{equation}
where $r_{0}$ denotes the deviation from the critical point $r_{0} = a(\mu-\mu_{c})$ with a constant $a$, and $g$ denotes the coupling constant.
$V_{0}$ represents a contribution independent of $\psi$.
The critical point is at $\mu=\mu_{c}$ with a fixed temperature.
We also denote the spatial derivative $(\partial_{x},\partial_{y},\partial_{z})$ by $\vec{\nabla}$.
In the GL theory, the complex scalar field represents the order parameter.
Considering the time-independent case, we obtain the equation of motion for the order parameter:
\begin{equation}
    0 = \frac{\delta \Omega}{\delta \psi^{*}} = -\frac{1}{2}\vec{\nabla}^{2}\psi +\frac{1}{2}r_{0}\psi +\frac{1}{2}g \psi \left|\psi \right|^{2}.
\end{equation}
Below the critical point ($\mu<\mu_{c}$), we find a well-known kink solution
\begin{equation}
    \psi(x) = \Delta\tanh{\left( \frac{x}{\xi} \right)},
\end{equation}
where $\Delta = \sqrt{a(\mu_{c}-\mu)/g}$, and we impose the boundary conditions $\psi(x\to \infty)=\Delta$ and $\psi(x\to -\infty)= -\Delta$.
This solution interpolates between two potential minima $\psi = \pm \Delta$.
The healing length is given by
\begin{equation}
    \xi = \sqrt{\frac{2}{a(\mu_{c}-\mu)}}.
    \label{eq:GLscale}
\end{equation}
If we consider the charge density $\rho$ is given by the density of the order parameter, we obtain
\begin{equation}
    \frac{\rho}{\rho_{\rm homo}} \sim \frac{\left| \psi\right|^{2}}{\Delta^2} \sim  \tanh^{2} \left(\frac{x}{\xi}\right)= 1- \sech^{2} \left( \frac{x}{\xi} \right).
    \label{eq:GLcharge}
\end{equation}

\begin{figure}[tbp]
    \centering
    \includegraphics[width=7cm]{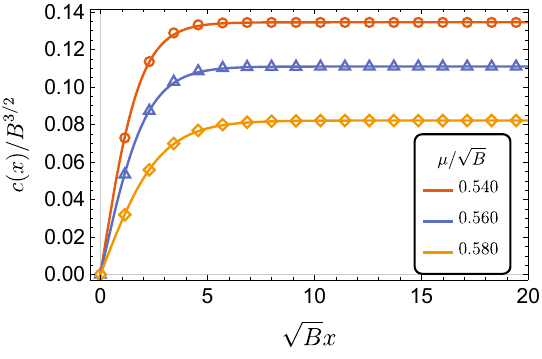}
    \includegraphics[width=7cm]{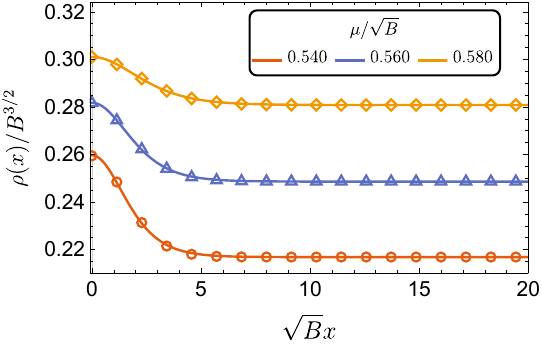}
    \caption{The fitting results of the chiral condensate (left) and charge density (right) profiles for several values of the chemical potential with $\pi T/\sqrt{2B}=0.070711$ fixed.
    The points denote the numerical data, and the curves denote the fitting results with (\ref{eq:fit}).
    }
    \label{fig:fit}
\end{figure}

\begin{figure}
    \centering
    \includegraphics[width=10cm]{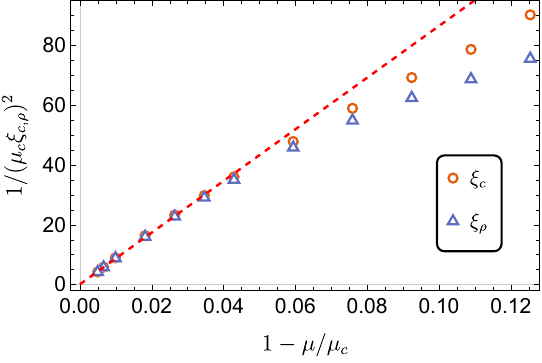}
    \caption{The relation between the healing lengths $(\xi_{c},\xi_{\rho})$ and the chemical potential. The red dashed line denotes the linear fitting result of the first few plots near the critical points.}
    \label{fig:xi}
\end{figure}

Motivated by this, we study the healing length by fitting the numerical data with the following fitting functions:
\begin{equation}
    c(x) = c_{\rm homo} \tanh (\frac{x}{\xi_{c}}), \quad
    \rho(x) = \rho_{\rm homo}
    - \left(
        \rho_{\rm homo} - \rho_{0}
    \right) \sech^{2}\left( \frac{x}{\xi_{\rho}}\right),
     \label{eq:fit}
\end{equation}
where $(\xi_{c},\xi_{\rho})$ are the two different healing lengths for the chiral condensate and charge density, respectively.
Note that we introduce an extra parameter $\rho_{0}$ which corresponds to $\rho(0)$, in order for the charge density function to describe our numerical results well.
We consider that the difference with (\ref{eq:GLcharge}) originates from the fact that the order parameter and charge density are related by more complicated way in our model, that is, they are obtained from the solutions of two coupled but independent fields.
Nevertheless, our fitting functions are compatible with the GL theory as demonstrated later.
Figure \ref{fig:fit} shows that the fitting of the chiral condensate and charge density for several values of $\mu/\sqrt{B}$ at $\pi T/\sqrt{2B} = 0.070711$.
As shown, the numerical profiles are well-fitted by the functions (\ref{eq:fit}) with the fitting parameters $(\xi_{c},\xi_{\rho}, \rho_{0})$.
Furthermore, we show the relation between the healing lengths obtained from the fitting and the chemical potential in figure \ref{fig:xi}.
We find that the healing lengths, $\xi_{c}$ and $\xi_{\rho}$, have different dependencies on the chemical potential.
However, these two length scales are coincident in the vicinity of the critical point and numerically given by 
\begin{equation}
    \frac{1}{\mu_{c}\xi} \approx 29.4 \sqrt{1- \frac{\mu}{\mu_{c}}},
\end{equation}
which agrees with the result from the GL theory (\ref{eq:GLscale}).
Since the GL theory is expected to be valid only near the critical point, our results imply that the kink solutions we found could be well-described by the GL theory in the vicinity of the critical point.

Moreover, we should note that this behavior is also compatible with the static perturbation scale (\ref{eq:staticpert}), relating to the inhomogeneous length scale, obtained from the linear perturbation analysis in section \ref{sec:static}.
Since the analysis is perturbatively performed on the solutions in the $\chi$SR phase, it should represent the behavior near the critical point.
Hence, we conclude that the healing length, which characterizes the length scale of the kink solution we obtained at the non-linear level, is indicated (up to constant factor) by the linear perturbation analysis in the vicinity of the critical point.

Before closing this subsection, we comment on the consistency between our fitting functions and the GL theory.
In figure \ref{fig:xi}, it is shown that $\xi_c$ and $\xi_{\rho}$ take almost same value near $\mu=\mu_c$.
This observation may be understood as follows.
Since the thermodynamic potential is almost described by Eq.~(\ref{eq:4th-GL}) near $\mu=\mu_c$ by replacing $\psi(x)$ with $c(x)/c_{0}$, where $c_{0}$ is a constant fixing the dimensionality, it is expected that the variation of Eq.~(\ref{eq:4th-GL}) with respect to $\mu$ gives the charge density.
In this case, $r_0$ has linear dependence on $\mu$ as $r_0=a(\mu-\mu_c)$, whereas $g$ does not.
Writing the local potential by $\Omega = \int\dd[3]{x}\Omega(x)$, we obtain the estimation for $\rho(x)$ as
\begin{equation}
    \rho(x) \sim
    - \left.\pdv{\Omega(x)}{\mu}\right|_{\psi(x)=c(x)/c_{0}}
    = - \frac{a}{2} \frac{c(x)^2}{c_{0}^2} - \pdv{V_0}{\mu}.
\end{equation}
Note that the second term $\pdv*{V_0}{\mu}$ does not depends on the spatial coordinate $x$.
This expression of the estimation agrees with our fitting function (\ref{eq:fit}), and $\xi_c$ and $\xi_{\rho}$ must be identical near the critical point.

\subsection{Near the first order phase transitions}
\begin{figure}
    \centering
    \includegraphics[width=8.0cm]{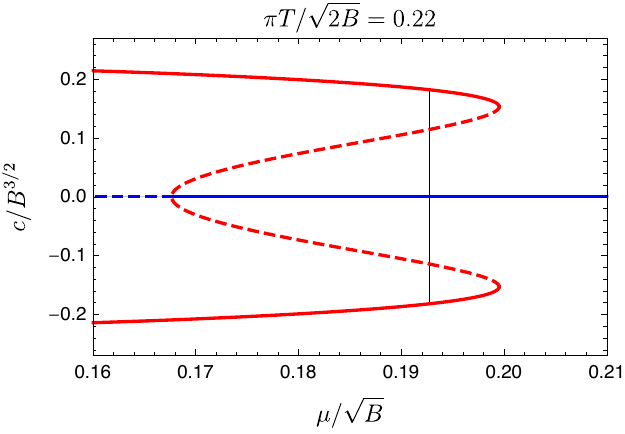}
    \caption{$\mu$--$c$ relation for the homogeneous setup at $\pi T/\sqrt{2B} = 0.220$.
    The red (dashed) curve denotes the (unstable) $\chi$SB branch.
    The blue (dashed) line denotes the (unstable) $\chi$SR branch.
    The vertical black line indicates the location of the first order phase transition point at $\mu=\mu_{\rm 1st} = 0.19277 \sqrt{B}$.}
    \label{fig:mu-c_1stPT}
\end{figure}%
So far we have investigated the kink solutions near the second order phase transition points.
Now let us study their profiles near the first order phase transition points.
In the case of the homogeneous setup, we can find two different branches of solutions in the chiral symmetry broken phase with a given chemical potential.
Figure \ref{fig:mu-c_1stPT} shows the relation between $c$ and $\mu$ at $\pi T/\sqrt{2B}=0.220$.
In the $\chi$SB branch, one solution is thermodynamically stable (solid curve) and the other is unstable (dashed curve).
The first order phase transition undergoes between the solutions in the $\chi$SR and $\chi$SB phases at a specific value of the chemical potential (vertical black line) when the temperature is fixed.

\begin{figure}[tbp]
    \centering
    \includegraphics[width=7cm]{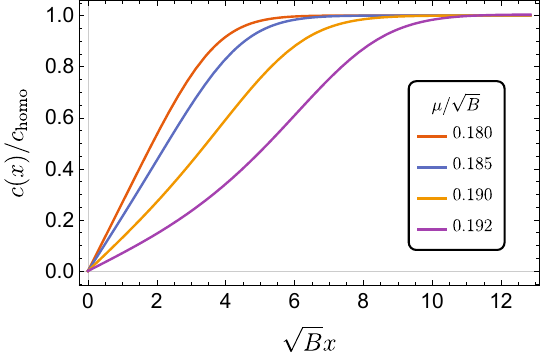}
    \includegraphics[width=7cm]{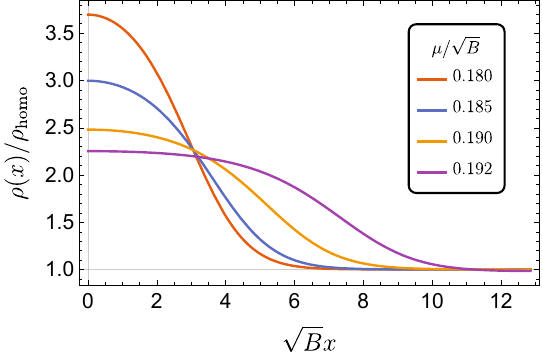}
    \caption{The profiles of $c(x)$ (left) and $\rho(x)$ (right) normalized by the homogeneous (stable-$\chi$SB) values $c_{\rm homo}$ and $\rho_{\rm homo}$ for several values of $\mu/\sqrt{B}$ with $\pi T/\sqrt{2B} = 0.220$ fixed, near the first order phase transition point.
    }
    \label{fig:cmrm1}
\end{figure}

\begin{figure}[tbp]
    \centering
    \includegraphics[width=7cm]{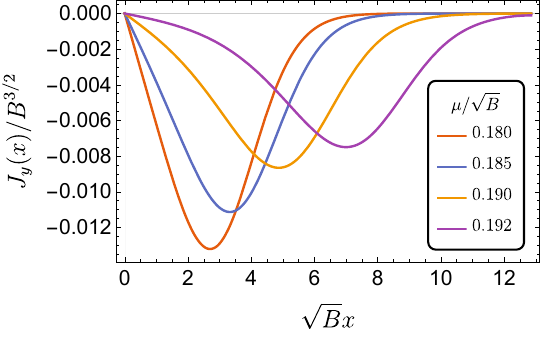}
    \includegraphics[width=7cm]{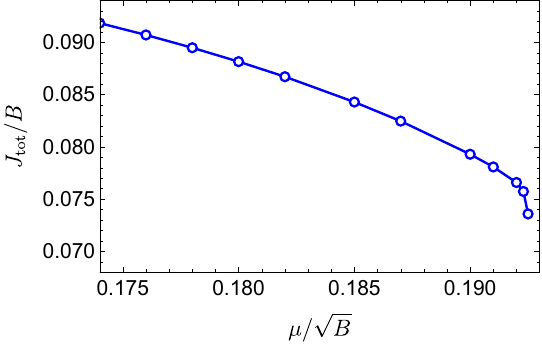}
    \caption{The profiles of $J_{y}(x)$ for several chemical potentials (left) and the chemical potential dependence of $J_{\rm tot}$ (right) with $\pi T/\sqrt{2B} = 0.220$ fixed, near the first order phase transition point.
    }
    \label{fig:jym1}
\end{figure}

In the inhomogeneous setup, on the other hand, we find only one branch of solutions which approach the homogeneous solution at the spatial infinity with a given homogeneous chemical potential.
We show the profiles of the chiral condensate and charge density for several values of the chemical potential in figure \ref{fig:cmrm1}.
We also show the profiles of the chiral edge current and the total current with respect to the chemical potential in figure \ref{fig:jym1}.
As shown the profiles deform in a qualitatively different way compared with those near the second order phase transitions.
Notably, these profiles do not continuously approach those in the $\chi$SR phase, unlike $\rho(x)$ shown in Fig.~\ref{fig:cmrm} near the second order phase transition.
As $\mu$ increases, the inhomogeneous solutions in the $\chi$SB phase are not found above a specific value of $\mu$.
This suggests a discontinuous transition between the $\chi$SB and $\chi$SR phases, corresponding the first order phase transition undergoes.
Additionally, the solutions are no longer well described by (\ref{eq:fit}) derived from the 4th order GL theory.
This fact can be natural because the 4th order GL theory generally describes the vicinity of the second order phase transition points, therefore we require to consider another theory.
One of the candidates describing the first order phase transitions is 6th order GL theory.
In the following subsection, we compare our numerical results with the analytic results derived from the 6th order GL theory.

\subsection{Comparison to the 6th order Ginzburg--Landau theory}
\begin{figure}[tbp]
    \centering
    \includegraphics[width=10cm]{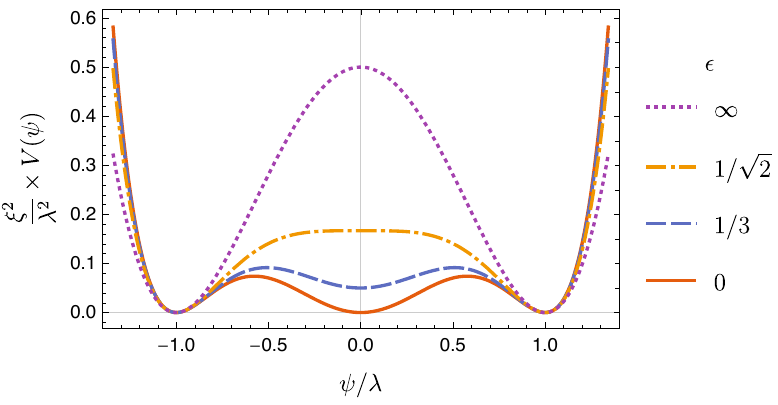}
    \caption{
    Potential of the 6th order GL theory for various $\epsilon$ taking value in $0\leq\epsilon\leq\infty$.
    In the limit of $\epsilon\to\infty$, the potential reduces to the 4th order.
    Note that we have scaled out $\lambda$ and $\xi$.
    We have also set $V_0 = 0$ and $\psi$ as a real number here.
    }
    \label{fig:potential}
\end{figure}%
To compare our solutions near the first order phase transition, we consider the 6th order GL theory.
See \cite{Saxena:2018axk} for further details of this model.
The thermodynamic potential is given by Eq.~(\ref{eq:4th-GL}) with
\begin{equation}\label{eq:6th-GL}
    V(\psi)=
    \frac{1}{2\xi^2\lambda^4(1+\epsilon^2)}
    (\abs{\psi}^2+\epsilon^2\lambda^2)
    (\abs{\psi}^2-\lambda^2)^2 + V_0,
\end{equation}
where $\lambda$, $\xi$ and $\epsilon$ are positive parameters.
Figure \ref{fig:potential} shows the shapes of the potential for various $\epsilon$.
If $\epsilon<1/\sqrt{2}$, the potential has three local minima at $\psi=\pm\lambda$ and $0$.
Due to the choice of the parametrization, this model only covers cases that $V(\pm\lambda)\leq V(0)$.
The energy difference between two metastable vacua is given by
$\Delta V = V(0) - V(\pm\lambda) = \epsilon^2 \lambda^2/(2(1+\epsilon^2)\xi^2)$.
Taking $\epsilon\to\infty$, the model reduces to the standard 4th order GL theory.
The three vacua are degenerate when $\epsilon=0$, and this case must correspond to the first order phase transition point.
Therefore, we expect that the energy difference $\Delta V$ is proportional to $\mu_{\text{1st}} - \mu$ near the phase transition point:
\begin{equation}
    \mu_{\text{1st}} - \mu
    \propto
    \frac{\lambda^2}{\xi^2} \epsilon^2.
    \quad (\text{for}~\mu \simeq \mu_{\text{1st}})
    \label{eq:epsilonscale}
\end{equation}
Note that $\xi$ and $\lambda$ also depend on $\mu$, but only $\epsilon$ will be relevant to the first order phase transition point.

This model has the (anti)kink solution called {\it bound pair},
\begin{equation}
    \psi(x) = \pm \lambda \frac{\sinh(x/\xi)}{\sqrt{1+\epsilon^{-2}+\sinh^{2}(x/\xi)}}.
    \label{eq:kink_bound-pair}
\end{equation}
This solution interpolates between $\psi(-\infty) = \mp\lambda$ and $\psi(\infty)=\pm\lambda$, respectively.
Taking $\epsilon\to\infty$, the kink solution reduces to $\psi(x)=\lambda \tanh(x/\xi)$.
At the first order phase transition point, corresponding to $\epsilon =0$, the following half (anti)kinks become solutions instead of the kink solutions (\ref{eq:kink_bound-pair}):
\begin{equation}\label{eq:half-kink}
   \psi(x) = \pm \lambda \sqrt{\frac{1\pm \tanh(x/\xi)}{2}},\quad
   (\text{for}~\epsilon=0)
\end{equation}
where two signs can be chosen independently so there are four solutions.
These solutions interpolate between $\psi(-\infty)=0$ (or $\pm\lambda$) and $\psi(\infty)=\pm\lambda$ (or $0$).

\begin{figure}[tbp]
    \centering
    \includegraphics[width=7cm]{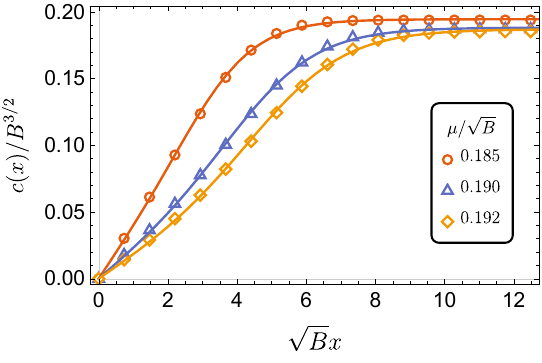}
    \includegraphics[width=7cm]{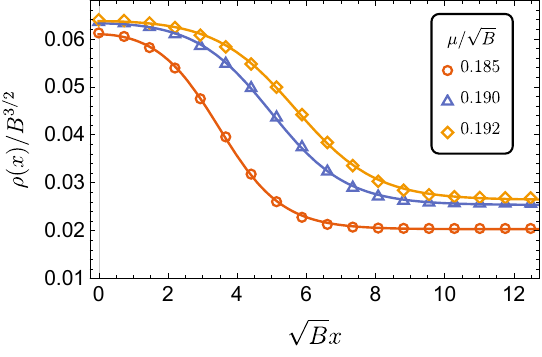}
    \caption{The fitting results of the chiral condensate (left) and charge density (right) profiles for several values of the chemical potential with $\pi T/\sqrt{2B}=0.22$ fixed, near the first order phase transition point.
    The points denote the numerical data, and the curves denote the fitting results with (\ref{eq:fit1}).
    }
    \label{fig:fit1st}
\end{figure}

Let us now examine how the 6th order GL theory agrees with our solutions near the first order phase transition points.
The fitting functions for the chiral condensate and charge density are respectively
\begin{equation}
    c(x) =  c_{\rm homo} \frac{\sinh(x/\xi_{c})}{\sqrt{1+\epsilon^{-2}_{c}+\sinh^{2}(x/\xi_{c})}}, \quad
    \rho(x) = \rho_{0} - (\rho_0 - \rho_{\rm homo}) \frac{\sinh^{2}(x/\xi_{\rho})}{1+\epsilon^{-2}_{\rho}+\sinh^{2}(x/\xi_{\rho})},
    \label{eq:fit1}
\end{equation}
where we introduce the fitting parameters $(\xi_{c,\rho},\epsilon_{c,\rho})$ separately as in section \ref{sec:4thGL}.
We also introduce an extra parameter $\rho_{0}$ in the same manner as in (\ref{eq:fit}).
In figure \ref{fig:fit1st}, we show the fitting results for several values of $\mu$.
It turns out that our numerical solutions near the first order phase transition point are well-fitted to the bound pair solutions.

\begin{figure}[tbp]
    \centering
    \includegraphics[width=7cm]{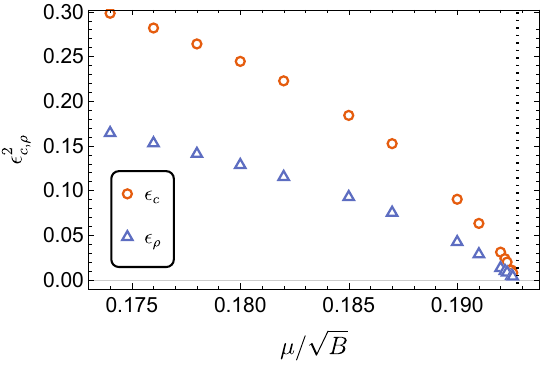}
    \includegraphics[width=7cm]{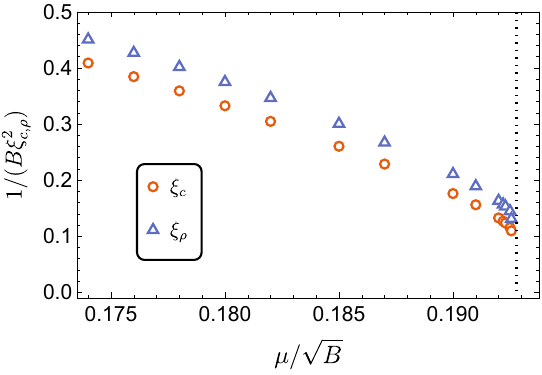}
    \caption{The behaviors of $\epsilon_{c,\rho}$ (left) and $\xi_{c,\rho}$ (right) with respect to the chemical potential.
    The vertical black dotted line indicates $\mu=\mu_{0}\approx 0.19277\sqrt{B}$ determined by the fitting.}
    \label{fig:epsilonxi}
\end{figure}

\begin{figure}[tbp]
    \centering
    \includegraphics[width=7cm]{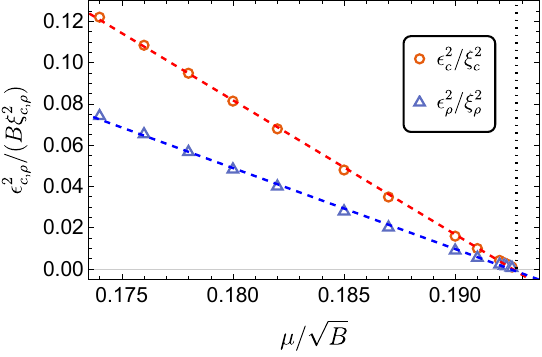}
    \caption{
        $\epsilon_{c,\rho}^2/(B \xi_{c,\rho}^2)$ as a function of $\mu/\sqrt{B}$.
        The red and blue dashed lines denote the linear fitting in $\mu_{0}-\mu$.
        The vertical black dotted line indicates $\mu=\mu_{0}\approx 0.19277\sqrt{B}$.
    }
    \label{fig:e-x1st}
\end{figure}

Then, we show the fitting parameters $(\xi_{c,\rho},\epsilon_{c,\rho})$ as functions of the chemical potential in figure \ref{fig:epsilonxi}.
For larger $\mu$, we find that $\epsilon_{c,\rho}^{2}$ approaches zero at $\mu = \mu_{0} \approx 0.19277 \sqrt{B}$ in the both cases of $c,\rho$, which coincides with the first order phase transition point $\mu_{\rm 1st}=0.19277\sqrt{B}$ computed in the homogeneous setup.%
\footnote{%
In this paper, we determine the first order phase transition point, $\mu_{\rm 1st}$, in the homogeneous setup from the onshell DBI action with the same numerical scheme as the inhomogeneous one for consistency.
}
Thus, it could be argued that $\epsilon$ goes to zero as we approach the first order phase transition point, which agrees with the behaviors of the 6th order GL theory.
We also find that $1/\xi_{c,\rho}^2$ decreases as $\mu$ increases, but it appears to remain finite at $\mu=\mu_{0}$.
Figure \ref{fig:e-x1st} shows $\epsilon_{c,\rho}^2/(\xi^2_{c,\rho})$, which is closer to Eq.~(\ref{eq:epsilonscale}), as a function of $\mu/\sqrt{B}$.
It shows a clearer linear dependence on $\mu_{0} - \mu$ in both cases of $c$ and $\rho$.
This observation agrees with the expected behavior in the 6th order GL theory.


At the first order phase transition point, we expect that the half-kink solution is realized, which can be fitted by using Eq.~(\ref{eq:half-kink}).
Unfortunately, we cannot find such a solution in our model.
The reason why we cannot obtain the half-kink solution may be because the half kink is only realized at the first order phase transition point, and it requires numerically fine tuning of the parameters in the model.

Another possibility is that the size of the kink solution exceeds the size of the system $l=8$ we set. The right panel of figure \ref{fig:epsilonxi} implies that $\xi$ of the bound pair increases as we approach the first order phase transition point.
If the half-kink solution has a much larger $\xi$ than $l$, we can not capture such a solution within the current computational setup.
To investigate the kink solution just at the first order phase transition point correctly, we need more careful treatment of the computation.

\subsection{Thermodynamic potential} \label{sec:TP}
\begin{figure}[tbp]
    \centering
    \includegraphics[width=7cm]{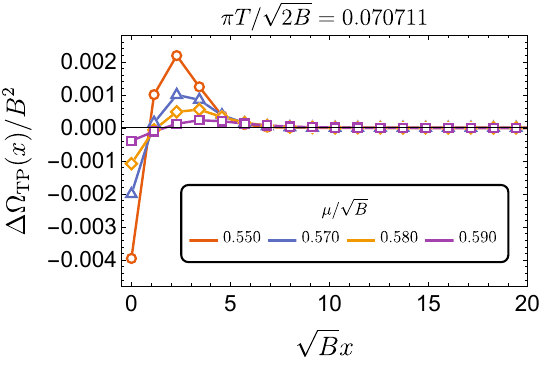}
    \includegraphics[width=7cm]{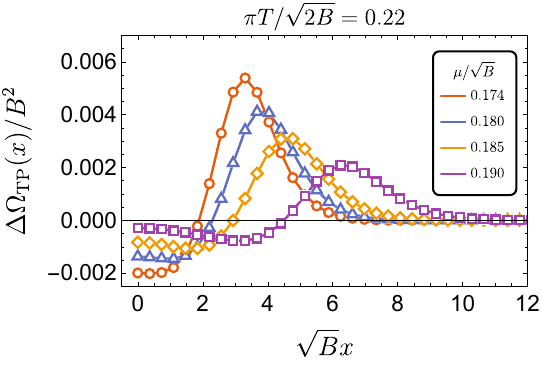}
    \caption{The spatial profiles of the thermodynamic potential difference near the second (left) and first (right) order phase transition points for several values of $\mu/\sqrt{B}$. The black dotted and red dashed lines denote those of the homogeneous solutions in the $\chi$SR and $\chi$SB phases, respectively. 
    }
    \label{fig:x-TP}
\end{figure}

As the last part of this section, we discuss thermodynamic stability of the kink solutions based on the thermodynamic potential.
The thermodynamic potential can be evaluated from the on-shell bulk action.
In our case with the inhomogeneous solutions, the $x$-dependent thermodynamic potential density may be computed as
\begin{equation}
    \Omega_{\text{TP}}(x) \equiv
    -\lim_{\varepsilon\to0}
    \left[
        \int_{\varepsilon}^{u_{\text{H}}}\dd{u}
        \left.\mathcal{L}(u,x)\right|_{\text{on-shell}}
        + L_{\text{ct}}^{\varepsilon}
    \right],
\end{equation}
where $\mathcal{L}$ is a Lagrangian density defined by $S_{\rm D7} = T_{\rm D7}(2\pi^2)\int\dd[4]{x}\dd{u}\mathcal{L}$ for the action (\ref{eq:DBI}), $\varepsilon$ is a small cutoff, and $L^{\varepsilon}_{\text{ct}}$ is an appropriate counterterm to subtract the divergence in the integral, as discussed in \cite{Karch:2007pd,Karch:2005ms}.
However, since we will show only the energy difference of the potential in the following, the detail of the counterterm does not affect the results.
Since the kink solutions approach the homogeneous solutions in the (stable) $\chi$SB phase, rather than in the $\chi$SR phase, at the spatial infinity, it is convenient to consider the following quantity:
\begin{equation}\label{eq:kink's_energy}
    \Delta\Omega_{\rm TP}(x) \equiv 
        \Omega_{\rm TP}^{\rm kink}(x)
        - \Omega_{\rm TP}^{\chi {\rm SB}}, 
\end{equation}
which is the difference of the thermodynamic potential between the kink and homogeneous solutions in the $\chi$SB phase.
Figure \ref{fig:x-TP} shows the $x$-dependent thermodynamic potential density of the kink solutions, and the corresponding homogeneous values near the second and first order phase transition points. 
As shown, the thermodynamic potential densities of the kink solutions approach those of the homogeneous solutions at $x=l/2$ as well as the profiles of the solutions.
We find that the thermodynamic potential becomes partly larger (or smaller) along with the inhomogeneity compared to that of the homogeneous solutions.
Note that the values of the potential of the kink at $x=0$ are different from those of the unstable $\chi$SR solutions, whose values are higher than $\Omega_{\rm TP}^{\chi\rm SB}$ in these parameter regimes.
On the other hand, the potential at $x=0$ of the kink approaches the same value as the potential of the stable $\chi$SB solution near the first order phase transition point, as shown in the right panel of Fig.~\ref{fig:x-TP}.
This behavior is consistent with the expectation that the half-kink solution (\ref{eq:half-kink}), which interpolates between the $\chi$SR and $\chi$SB solutions having the same potential energy, can be realized at the first order phase transition point.

\begin{figure}[tbp]
    \centering
    \includegraphics[width=7cm]{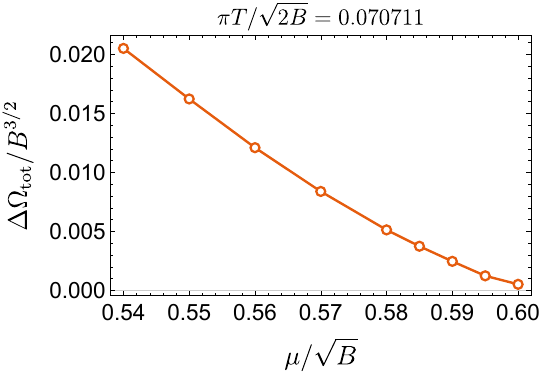}
    \includegraphics[width=7cm]{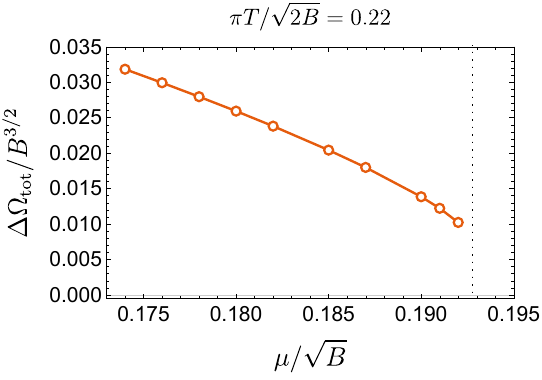}
    \caption{The difference of the thermodynamic potential between the kink and homogeneous solutions in the $\chi$SB phase near the second (left) and first (right) phase transition points.
    The vertical black dotted line denotes the first order phase transition point, $\mu_{\rm 1st} = 0.19277\sqrt{B}$.
    }
    \label{fig:mu-TP}
\end{figure}

Let us compare the total thermodynamic potential of the kink solutions with that of homogeneous ones.
Integrating (\ref{eq:kink's_energy}) over $x$, we write
\begin{equation}\label{eq:kink's_energy_total}
    \Delta\Omega_{\rm tot} = \int_{-l/2}^{l/2}\dd{x}\left[
        \Omega_{\rm TP}^{\rm kink}(x)
        - \Omega_{\rm TP}^{\chi {\rm SB}}
    \right].
\end{equation}
Note that $\Delta\Omega_{\rm tot}$ has the same dimension to the energy density on the remaining two-dimensional spatial coordinates ($y,z$).
From the results in figure \ref{fig:x-TP}, we can expect this quantity converges to a specific energy value associated with the kink solution regardless of the spatial cutoff $l$.
Figure \ref{fig:mu-TP} shows $\Delta \Omega_{\rm tot}$ as a function of $\mu$ near the second and the first order phase transition points.
As far as we have checked, the total difference of the thermodynamic potential between the kink solution and the homogeneous $\chi$SB solution, $\Delta \Omega_{\rm TP}$, is always positive in those parameter regions.

The analysis of the thermodynamic potential reveals that the homogeneous solutions are thermodynamically preferable to the kink solutions.
It also implies that a pair of the kink and anti-kink state will not be spontaneously produced from the homogeneous vacuum.
Still, there are possibilities that kink solutions can be realized because they are topologically protected and robust against local fluctuations. 
We can say that the kink solutions are metastable in this sense. 
In other words, a large energy would be necessary for the transition from the kink solutions to the homogeneous ones since the half of the configuration must be flipped or the kink needs to be transferred to the spatial infinity. 
Hence, the kink solutions can be formed such as in relaxation process or transient dynamics.

\section{Conclusion and Discussions} \label{sec:discussion}
In this paper, we find spatially inhomogeneous solutions, specifically a kink solution, in the D3/D7 model with a finite magnetic field, chemical potential, and temperature.
These solutions are found in the chiral symmetry breaking phase in the phase diagram studied in the homogeneous setup \cite{Evans:2010iy}.
We find that the electric current perpendicular to the magnetic field emerges with inhomogeneity, corresponding to the chiral edge current induced by the Lorentz force acting on the charged particles.
In this sense, the kink plays a role of an edge or defect on space.

We investigate the deformation of the profiles and the total amplitude of the chiral edge current with respect to the chemical potential and temperature.
We find that those behaviors have a qualitative distinction near the second and first order phase transition point. 
Close to the second order phase transition point, the profiles of inhomogeneous solutions continuously approach the flat solutions, corresponding to the chiral symmetry restored phase, for large chemical potentials and temperatures.
Additionally, we find that the profiles of the order parameter and charge density for the kink solution are well-fitted by the analytical solutions derived from the 4th order GL theory.
The healing lengths obtained from both profiles have the same scaling with respect to the chemical potential near the critical point.
In the vicinity of the critical point, the dependence of the healing length on the chemical potential agrees with the results from both the GL theory and the linear perturbation analysis.
Around the first order phase transition point, on the other hand, we find that their profiles are no longer given by the analytical solutions in the 4th order GL theory, but can be described by the 6th order theory.
It is natural because the first order phase transition can be described by the energy competition of the 6th order potential.
Our results strongly imply that the low-energy effective theory in our setup can be described by the 4th or 6th order GL theory about the chiral condensate partially.
On the other hand, the interaction between the chiral condensate and the $U(1)$ current operators are still nontrivial.
We also discuss the thermodynamic stability of the kink solutions based on the thermodynamic potential.
Comparing to the thermodynamic potential of the homogeneous solutions, we reveal that they are thermodynamically preferable to the kink solutions both near the second and first order phase transition points.
Although the kink solutions are not thermodynamically preferred as far as we have observed, there are possibilities of realization of these states thanks to their topological stability.

Now we make some remarks on our results.
The chiral edge current we observe is originated from the cyclotron motions of the charged particles induced by the magnetic field and the inhomogeneity of the charge density near the domain wall.
This implies that the kink plays a role of edge with respect to the three dimensional bulk region.
In this sense, it would be interesting to consider the relation to the bulk-edge correspondence.
Although the chiral edge current in our study is not quantized unlike the quantum Hall effect, the edge state is clearly reflected by the properties of bulk region.

We make a comment on a similarity between our observation for the chiral edge current and the Hall current. 
This can be roughly understood as follows. As shown in figure \ref{fig:bdry}, the charge density is localized around $x=0$. The similar density profile appears for the normal metal with the potential well around $x=0$. Thus it is natural that the Hall effect in the presence of the potential well can be observed in our setup. In the following, we will explain the detailed analysis.
In our results, the chiral edge current is proportional to the spatial variation of the charge density as Eq.~(\ref{eq:chiral_edge_current}).
One can see that Eq.~(\ref{eq:chiral_edge_current}) is similar to the expression of the Hall current in the D3-D7 model which is also dissipationless.
It was studied in \cite{OBannon:2007cex}, and \cite{Ammon:2009jt} for general electromagnetic fields.
If we consider applying a small electric field along the $x$-direction in homogeneous states, the (linear) Hall conductivity is given by
\begin{equation}
    \sigma_{yx}^{\rm Hall} = - \frac{B\rho}{B^2+\pi^4 T^4}.
\end{equation}
Note that $\sigma_{xy}^{\rm Hall} = -\sigma_{yx}^{\rm Hall}$.
In the case of figure \ref{fig:bdry}, the Hall conductivity is given by $\sigma_{yx}^{\rm Hall}/\sqrt{B} = -0.2372$ for the choice of $\rho = \rho(x=l/2)$, and $\sigma_{yx}^{\rm Hall}/\sqrt{B} = -0.2718$ for $\rho=\rho(0)$.
We may write
\begin{equation}
    J_y(x) = \sigma_{yx}^{\rm CE} \times (-1) B^{-1} \partial_x \rho(x),
\end{equation}
where $\sigma_{yx}^{\rm CE}$ is a scaling $1$ coefficient.
Remark that there is ambiguity for matching the scaling dimension; we used $B$ but one can choose other quantities, such as $\mu, T$.
The minus sign is inserted to match the sign of the coefficient with the Hall conductivity.
In the case of figure \ref{fig:bdry}, we have obtained $\sigma_{yx}^{\rm CE}/\sqrt{B} = -0.3426$.
The spatial variation of the charge density may play a similar role to the electric field along the $x$-direction, but it does not induce the current density along the $x$-direction, and the corresponding conductivity is different from the Hall conductivity in our model.

In this study, we restricted ourselves to the real-valued order parameters.
It is straightforward to extend them to the complex quantity and investigate complex kink solutions. To do so, we also need to consider the inhomogeneous profiles of the pseudo-scalar field $\psi$ which is coupled to the other fields via the Wess--Zumino term.
It would be interesting to study how the chiral edge current is changed in the complex kink solutions.
In addition, if we consider the condensation of the complex scalar field, vortex solutions are also expected to be observed even in our model.
We will address these issues in a future work.

Another interest is the stability of the inhomogeneous solutions.
In the holographic superfluid model, the dynamical instability for the kink solutions under transverse perturbations, the so-called snake instability, was studied in \cite{Guo:2018mip,Xu:2019msl}.
Our kink solutions might exhibit the snake instability and be dynamically unstable.
However, the kink solutions in our model always accompanies the chiral edge current in both side of the kink and they would possibly stabilize the kink structure under transverse perturbations.
This dynamical stability should be addressed in the future.

Additionally, an important application of the probe brane model is the realization of steady states.
In \cite{Imaizumi:2019byu,Matsumoto:2022nqu}, the authors found that the spontaneous chiral symmetry breaking is observed in the steady state with a constant current in the D3/D7 model.
In this manner, we expect that spatially inhomogeneous solutions could be appeared also in the steady state.
If those novel phases could exist, it is of significant interest from the viewpoint of condensed matter physics.

Lastly, we comment on the Nambu--Goldstone modes associated with the spontaneous translation symmetry breaking.
As the emergence of a kink breaks the one-dimensional spatial translation symmetry, a fluctuation of the kink position provides a gapless mode, namely the Nambu--Goldstone (NG) mode.
The NG mode associated with the translation symmetry breaking in our model is expected to behave as a diffusive mode rather than propagating mode because the system is open, as discussed in \cite{Fujii:2021nwp} (see also \cite{Minami:2015uzo,Hidaka:2019irz,Hongo:2019qhi}).
In fact, the NG mode associated with the chiral symmetry breaking in the D3/D7 model behaves as a diffusive mode as discussed in \cite{Ishigaki:2020vtr}.
It would be interesting to investigate the behavior of the NG mode associated with the translation symmetry breaking in our kink solutions, but we leave it for future work.

\section*{Acknowledgement}
S.~I.~is supported by NSFC China (No.\,12147158), and Shanghai Post-doctoral Excellence Program (No.\,2022245). 
R.~Y.~is partially supported by a Grant-in-Aid of MEXT for Scientific Research (KAKENHI Grant Nos.\,19K14616 and 20H01838).

\appendix
\section{Numerical method} \label{sec:numeric}
In this appendix, we explain the details of the numerical calculations.
The equations of motion we have to solve are the partial nonlinear differential equations. 
Since it is generally difficult to find analytical solutions, we numerically find the spatially inhomogeneous solutions. In this paper, we employ the Newton-Raphson method. Firstly, we prepare the $M\times N$ grid points for the ($u,x$) coordinates. We approximate the derivatives with the finite-difference method, that is,
\begin{align}
    &\partial_{u} \Psi(u,x) \simeq \frac{\Psi(u_{i+1},x_{j})-\Psi(u_{i-1},x_{j})}{2 \Delta u}, \quad \partial_{x} \Psi(u,x) \simeq \frac{\Psi(u_{i},x_{j+1})-\Psi(u_{i},x_{j-1})}{2 \Delta x}, \nonumber \\
    &\partial_{u}^{2} \Psi(u,x) \simeq \frac{\Psi(u_{i-1},x_{j})-2\Psi(u_{i},x_{j})+\Psi(u_{i+1},x_{j})}{\Delta u^{2}}, \nonumber \\
    &\partial_{x}^{2} \Psi(u,x) \simeq \frac{\Psi(u_{i},x_{j-1})-2\Psi(u_{i},x_{j})+\Psi(u_{i},x_{j+1})}{\Delta x^{2}},
\end{align}
where $\Psi=(\theta, a_{t},a_{y})$ and ($u_{i},x_{j}$) denotes the $(i,j)$-th grid point ($i=1,\cdots, M$ and $j=1,\cdots N$). $\Delta u$ and $\Delta x$ are the grid spacing for each coordinate. Here, we set $u_{1}=0$, $u_{M} = u_{\rm H}$, $x_{1}=-l/2$, and $x_{N}=l/2$ as the boundary for each coordinate. Hence, the system is given by the box whose size is $(0,1) \times l$. 
We fix $l=8$ in our actual calculations.
Since the scaled size $l\sqrt{B}$ is sufficiently large compared to the other scaled parameters ($\mu/\sqrt{B}, \pi T/\sqrt{B}$), we can regard it as a large spatial cutoff.
Actually, we have checked our results do not change qualitatively even if we vary $l$.
Then, we discretize the equations of motion for $\Psi(u,x)$ and obtain $(M-2)\times (N-2)$ equations for each field except for the boundary equations. 
Besides, the boundary conditions are given as follows.
At the black hole horizon ($i=1$), we impose the regularity conditions obtained from the equations of motion.
At the boundary cutoff in the $u$-coordinate ($i=M$), we impose the boundary condition (\ref{eq:asym1})-(\ref{eq:asym3}) as explained in section \ref{sec:setup}.
At the boundary in the $x$-coordinate ($j=1,N$), we impose the Neumann boundary condition.
As a result, we obtain the $3MN$ nonlinear equations for the fields $\Psi(u_{i},x_{j})$. Here, let us denote them as 
\begin{equation}
    E_{\alpha}\left( \Psi_{\beta} \right)=0,
    \label{eq:EE}
\end{equation}
where $\alpha=1,\cdots 3MN$ and $\Psi_{\beta} = \left\{ \theta(u_{1},x_{1}) , \cdots ,\theta(u_{1},x_{N}),\theta(u_{2},x_{1}), \cdots  \theta(u_{M},x_{N}), a_{t}(u_{1},x_{1}) , \cdots\right\}$ with $\beta =1,\cdots, 3MN$.
Assuming that $\bar{\Psi}_{\beta}$ are some configurations and (\ref{eq:EE}) is satisfied after we give a small deviation $\delta \Psi_{\beta}$, we can write
\begin{align}
    &E_{\alpha}(\bar{\Psi}_{\beta}+\delta \Psi_{\beta}) \simeq E_{\alpha} (\bar{\Psi}_{\beta})+ \delta \Psi_{\beta} \frac{\partial E_{\alpha}}{\partial \Psi_{\beta}} =0, \nonumber\\\
    & \Longrightarrow \delta \Psi_{\beta} \simeq - \left(J^{-1}\right)_{\beta}{}^{\alpha} E_{\alpha}(\bar{\Psi}_{\beta}),
\end{align}
where $J^{\beta}{}_{\alpha}\equiv \partial E_{\alpha} / \partial \Psi_{\beta}$ is the Jacobian matrix. 
In this way, we can obtain the deviation from the initial configuration $\bar{\Psi}_{\beta}$ to the true solution at the linear level.
Since the equations are nonlinear in $\Psi_{\beta}$, we iterate the computation of $\delta \Psi_{\beta}$ until it relaxes to zero. 
In the actual calculations, we use $(M,N)=(35,70)$ and iterate the calculations up until the mean absolute error becomes efficiently small: $\sum_{\beta} \left| \delta \Psi_{\beta} \right| / (3NM) \lesssim 10^{-10}$. 
We also confirm that the numerically obtained solutions does not drastically change by varying the values of $(M,N)$.

\bibliographystyle{ytphys}
\bibliography{main}
\end{document}